\documentclass[preprint,authoryear,5p]{elsarticle}  

\usepackage{placeins}
\usepackage{graphicx}	
\usepackage{amsmath}	
\usepackage{amssymb}	
\usepackage{hyperref}
\usepackage{natbib} 
\usepackage{lineno} 
\usepackage{adjustbox}

\hypersetup{draft}

\title{Non-principal axis rotation in binary asteroid systems and how it weakens the BYORP effect}

\author[a1]{Alice C. Quillen\corref{cor1}}
\ead{alice.quillen@rochester.edu}
\author[a1]{Anthony LaBarca}
\ead{alabarca@u.rochester.edu} 
\author[a2]{YuanYuan Chen}
\ead{chenyy@pmo.ac.cn}

\address[a1]{Department of Physics and Astronomy, University of Rochester, Rochester, NY 14627, USA}
\address[a2]{Key Laboratory of Planetary Sciences, Purple Mountain Observatory, Chinese Academy of Sciences, Nanjing 210008, China}

\cortext[cor1]{Corresponding author}

\begin{document}

\begin{abstract}
Using viscoelastic mass/spring model simulations, we explore tidal evolution and migration of compact binary asteroid systems.  We find that after the secondary is captured into a spin-synchronous state,  non-principal axis rotation in the secondary can be long-lived.    
The secondary's long axis can remain approximately aligned along the vector connecting secondary to primary while the secondary  rocks back  and forth about its long axis. 
Inward orbital semi-major axis migration can also  resonantly excite non-principal axis rotation.   
By estimating solar radiation forces on triangular surface meshes, we show that the  magnitude of the BYORP effect induced torque is sensitive to the secondary's spin state.
Non-principal axis rotation within the 1:1 spin-orbit resonance can reduce  the BYORP torque
or cause frequent reversals in its direction.  
\end{abstract}

\maketitle

\section{Introduction}
\label{sec:intro}

The field of research on binary asteroids currently has two dynamical conundrums.  
Light curve based estimates of viscoelastic behavior in single asteroids based on tumbling damping give elastic modulus times energy dissipation parameter, $\mu Q \sim 10^{11}$ Pa \citep{harris94,Pravec_2014}, exceeding those estimated in binary asteroids \citep{goldreich09,taylor11,Jacobson_2011,Scheirich_2015,Nimmo_2019} by 1 to 4 orders of magnitude.
Drift rates estimated due to radiation forces, known as binary-YORP or BYORP, are so rapid that binary NEA asteroid lifetimes should be short, $\lesssim 10^5$ years \citep{Cuk_2005}. This short lifetime makes it difficult to account for the abundance of binary NEA asteroids, about one sixth of NEAs,  \citep{Margot_2002,Pravec_2006} within the NEA asteroid orbital lifetime, $ \sim 10^7$ years \citep{Gladman_2000}.

With asteroid binary light curves,   \citet{Pravec_2016} find that synchronous secondary rotation predominates in close binaries and in binaries with near circular orbits.    Only binary members that occupy a spin-orbit resonance, such as the synchronous 1:1 state, are expected to have non-zero cumulative BYORP effects. Otherwise the time-averaged torques are expected to cancel out \citep{Cuk_2005}.  
The rapid predicted BYORP drift rate, of order a few cm/yr \citep{Cuk_2005},  spurred \cite{Cuk_2010} to explore scenarios where the tidal torque is neglected. In contrast,  \citet{Jacobson_2011} proposed that BYORP induced orbital drift could be halted by tidal torque with sufficient tidal dissipation.  
Orbital eccentricity could  cause an elongated  secondary to tumble, 
delaying BYORP drift-induced  evolution and extending the lifetime of NEA binaries \citep{Naidu_2015}.

Possible ways to account for the spin synchronous rotation and the inferred tidal spin-down of the secondary are earlier formation in the Main Belt, formation at a slower spin rate, formation at a smaller orbital semi-major axis where the spin down time is shorter, or if the secondary has lower $\mu Q \sim 10^9$ Pa \citep{taylor11}.  Rubble pile NEAs may have low $\mu Q $ values due to stress on contact points arising from internal voids \citep{goldreich09}, a dissipative regolith surface layer $\mu Q \sim 10^8$ Pa  \citep{Nimmo_2019} or because the rheology is better described with a frequency dependent model \citep{efroimsky15}.   These studies have neglected the possible role of some processes that could affect the spin down rate, such as mass shedding from the primary \citep{Walsh_2008,Hirabayashi_2015,Scheeres_2015} that could give a population of objects that might impact or accumulate on the secondary (e.g., \citealt{Davis_2020}),  ongoing mass loss, recently  detected on asteroid (101955) Bennu \citep{Lauretta_2019}, and mass movements as inferred from asteroid surface morphology \citep{Jawin_2020}.

With multiple epoch photometric observations spanning 20 years,  \citet{Scheirich_2015,Scheirich_2021} have directly measured binary orbital drift rates in 3 binary asteroid systems. In the case of (88710) 2001 SL9, the BYORP effect is the only known physical mechanism that can cause the measured $\sim -4$ cm/yr inward drift of its mutual binary orbit.  For NEA (66391) 1999 KW4, known as Moshup,  \citet{Scheirich_2021} measured a drift rate of 1.2 cm/s which is lower than that predicted for the BYORP effect, $\sim 8.5$ cm/s, using the secondary shape model by \citet{Ostro_2006}.

Apollo-class Near-Earth Asteroid (65803) Didymos is the target of the international collaboration known as AIDA (abbreviation for Asteroid Impact \& Deflection Assessment) that supports the development and data interpretation of the NASA DART  \citep{Cheng_2016} and ESA Hera \citep{Michel_2021} space missions.  AIDA associated missions will provide a unique opportunity to study a binary asteroid system. The secondary, called Dimorphos, of the Didymos system is the target of the Double Asteroid Redirection Test (DART) mission. The DART spacecraft will impact the binary's secondary fall 2022 to study the momentum transfer caused by a kinetic impact. The Light Italian Cubesat for Imaging of Asteroids (LICIA cube; \citealt{Dotto_2021}), carried with the DART spacecraft, will take in situ observations of the impact. Detailed physical surface and interior characterization as well as images of the surface morphology of the two bodies in the Didymos system by the Hera mission in early 2027 will probe how deformation occurs in a microgravity environment, giving constraints on binary asteroid internal structure, formation and evolution mechanisms.

\subsection{Outline}

Inspired by the opportunity represented by DART and associated missions, we examine the spin evolution of
the Didymos binary system. In section \ref{sec:Didymos} we review estimates for the tidal spin down time of both
bodies in the system.  
In section \ref{sec:sims} we explore rotation states and evolution of a binary asteroid system using our mass/spring model simulation code  \citep{quillen19_moon,quillen19_wobble,quillen20_phobos}.
Our simulations are similar to the soft sphere simulations described by \citet{Agrusa_2020}, except internal dissipation occurs due to damping in the springs.  Because we are interested in long term evolution, our simulations are carried out for more rotation periods than those by \citet{Agrusa_2020}.
In section \ref{sec:sim1} we use mass-spring model viscoelastic simulations  to  examine the process of secondary tidal spin down and tidal lock.  
In section \ref{sec:mig} we mimic  subsequent BYORP induced migration in the binary orbital semi-major axis $a_B$ through forced migration and examine its affect on the spin dynamics.
In section \ref{sec:BYORP} using triangular surface models we explore how obliquity and non-principal axis (NPA) rotation affects estimates of the BYORP induced drift rate.  
A summary and discussion follows in section \ref{sec:sum}.

\subsection{Properties of the Didymos binary system}
\label{sec:Didymos}

In Table \ref{tab:Didymos},  we list properties of the Didymos binary system.
We refer to the primary as Didymos and the secondary as Dimorphos.
Many of the quantities in this table were measured or recently reviewed by  \citet{Naidu_2020}.

\begin{table*}[] \centering
\caption{\Large Didymos and Dimorphos  properties}
\label{tab:Didymos}
\begin{adjustbox}{width=6.5in,center}
\begin{tabular}{llll}
\hline
Parameter & Symbol & Value & Comments/Reference \\
\hline
Binary orbital period & $P_B$ & 11.9217 $\pm $ 0.0002 hour & \citet{Scheirich_2009} \\
Binary orbital semi-major axis & $a_B$ &  1.19 $\pm$ 0.03 km & \citet{Naidu_2020} \\
Binary orbit eccentricity  & $e_B$ & $ < 0.03$ &  \citet{Scheirich_2009} \\
Didymos spin rotation period & $P_p$ & 2.2600 $\pm$ 0.0001 hours & \citet{Pravec_2006} \\
Didymos body diameter & $2 R_p$ & 780 $\pm$ 30  m  & \citet{Naidu_2020}\\
Diameter or radius ratio & $R_s/R_p$ & 0.21 $\pm$ 0.01 & \citet{Scheirich_2009} \\
Semi-major axis/radius primary &  $a_B/R_p$ & 3.05 & \\
Dimorphos diameter & $2 R_s $ & 164 $\pm $ 18 m & From $R_p$ and $R_s/R_p$ measurements\\
Dimorphos spin rotation period & $P_s$ & 11.92  hours & Spin synchronous state assumed \\
Total mass &  $M_p + M_s$ &  $5.37 \pm 0.44 \times 10^{11}$ kg & Kepler's third law \\
Mass ratio & $q=M_s/M_p$ & 0.0093  & From ratio of radii \\
Mass of Didymos & $M_p$ & $5.32 \times 10^{11}$ & From mass ratio \\
Mass of Dimorphos & $M_s$ & $4.95 \times 10^{9}$ & From mass ratio \\
Mean density of Didymos & $\rho_p$ & $2.14 \   {\rm g/cm}^{3} $ & From mass and radius \\
Mean density of Dimorphos & $\rho_s$ & $\rho_s = \rho_p$ & Assumed \\
Period ratio (orbit/spin primary) & $P_B/P_p$ & 5.28  & \\
Didymos body axis ratios & $a_p:b_p:c_p$ & 1:0.98:0.95 & Based on \citet{Naidu_2020} \\
%
\hline
\end{tabular} 
\end{adjustbox}
\\ \small
Radii $R_p,R_s$ for the primary and secondary are assumed to be those of the volume equivalent sphere. 
We refer to the primary as Didymos and the secondary as Dimorphos.
Dydimos' body axis ratios are based on the Dynamically Equivalent Equal Volume Ellipsoid which is an uniform density ellipsoid with the same volume and moments of inertia as the shape model by \citet{Naidu_2020} and taken from their Table 3.  Light curve observations by \citet{Pravec_2006} and radar data by \citet{Naidu_2020} are consistent with the secondary being in the spin synchronous state (tidally locked). The assumption of equal primary and secondary densities is supported by radar observations of NEA 2000 DP107 \citep{Margot_2002}.
\end{table*}

\begin{table*}[] \centering
\caption{Didymos and Dimorphos estimated properties}
\label{tab:estimated}
\begin{tabular}{lllc}
\hline
energy density & $e_{g,p}$ & 830  Pa \\
energy density & $e_{g,s}$ & 37  Pa \\
gravity timescale & $t_{\rm grav,p} $  & 1282 s \\
breakup spin & $\omega_{\rm breakup,p}$ & $ 7.8 \times 10^{-4} {\rm s}^{-1}$\\
\hline
\end{tabular}\\
\end{table*}


\subsection{Tidal spin down rates for primary and secondary}
\label{sec:spindown}

To estimate the tidal torque, 
we consider a uniform density spherical body that is part of binary system.  
The body is initially not tidally locked, so it is not in the spin synchronous state which is also called the 1:1 spin-orbit resonance. 
The spin-down rate $\dot \omega$ and a spin down time $t_{\rm despin}$ can be estimated from the tidally generated torque $\tau_{\rm tidal}$ with $\dot \omega_{\rm tidal} = \tau_{\rm tidal}/I$ and 
$t_{\rm despin} \sim \dot \omega_{\rm tidal}/\omega_{init}$
(e.g., \citealt{peale77,gladman96}) where   
the body's moment of inertia, $I = \frac{2}{5} MR^2$, the body's initial spin is $\omega_{\rm init}$ and the body's mass and radius are $M$ and $R$.
 
The secular part of the semi-diurnal ($l = 2$) term in the Fourier expansion of the perturbing potential from point mass $M_{pert}$ (the other mass in the binary),  gives a tidally induced torque 
(e.g., \citealt{kaula64,goldreich63,efroimsky13})
\begin{equation}
 \tau_{tidal} = \frac{3}{2} \frac{G M_{pert}^2}{a_B} \left( \frac{R}{a_B} \right)^5 k_{2} \sin \epsilon_2 \label{eqn:T}
\end{equation}
where orbital semi-major axis is $a_{B}$.
Here $k_2 \sin \epsilon_2$ depends on frequency and is known as a quality function. It is often approximated as $k_2/Q$ with energy dissipation or quality factor $Q$ 
describing the dissipation in the body with mass $M$, and the Love number $k_2$ characterizes the deformation of the body with mass $M$. 
See \citet{efroimsky15} for discussion on the accuracy of this approximation.

For an incompressible homogeneous spherical elastic body, the Love number 
\begin{equation}
k_2 \sim 0.038 \frac{e_g}{\mu} \label{eqn:k2}
\end{equation} 
where $\mu$ is the body's rigidity (equal to the shear modulus if the body is incompressible) and
\begin{equation}
e_g \equiv \frac{GM^2}{R^4} \label{eqn:eg}
\end{equation}
is a measure of the body's gravitational energy density,
(following Equation 6 by \citealt{quillen17_pluto} and based on \citealt{M+D,burns73}).
Note that at constant density, $M \propto R^3$, the energy density $e_g \propto R^2$
and Love number $k_2 \propto R^2$ which gives a size dependent Love number similar to (but not identical to) the weaker $k_2 \propto R$ proposed by \citet{goldreich09}. 

Using the torque of equation \ref{eqn:T}, 
and equation \ref{eqn:k2},
we estimate a spin down rate 
\begin{equation}
    \dot \omega_{tidal}
    \sim 0.14 \left( \frac{GM^2/R^4}{\mu Q}\right) \left(\frac{R}{a_B} \right)^6
    \left( \frac{ M_{pert}}{M} \right) 
    \frac{GM_{pert}}{R^3} .
\end{equation}

A particle resting on the surface can leave due to centrifugal acceleration if the spin 
$\omega \gtrsim \omega_{\rm breakup}$ with 
\begin{equation}
    \omega_{\rm breakup} \equiv \sqrt{\frac{GM}{R^3}}
    = \sqrt{\frac{ 4\pi G\rho}{3}} , \label{eqn:om_breakup}
\end{equation}
where $\rho$ is the body's mean density.
If the initial spin is equal to the that of centrifugal breakup $\omega_{\rm init} = \omega_{\rm breakup}$
then 
\begin{equation}
t_{\rm despin} \approx P_{orb} \frac{2 }{15\pi}
\left(\frac{M}{M_{pert}}\right)^\frac{3}{2}
\left(\frac{a_B}{R}\right)^\frac{9}{2} \frac{Q}{k_2} 
\sqrt{\frac{M_{pert} + M}{M_{pert}}} \label{eqn:t_despin}
\end{equation}
where $P_{orb} = {2 \pi} \sqrt{\frac{a_B^3}{G(M_{pert} + M)}}$ is the binary orbital period.

Using equations \ref{eqn:k2} and \ref{eqn:eg} the spin down time is 
\begin{equation}
\frac{t_{\rm despin}}{ P_B} \approx 1
\left(\frac{M}{M_{pert}}\right)^\frac{3}{2}
\left(\frac{a_B}{R}\right)^\frac{9}{2} \left(\frac{\mu Q}{GM^2/R^4}\right) \sqrt{\frac{M_{pert} + M}{M_{pert}}} .\label{eqn:t_despin2}
\end{equation}
This equation can be used for either a spinning primary or a spinning secondary.
In the case of a spinning primary 
\begin{align}
\frac{t_{\rm despin,p} }{ P_B}&\approx  
\left(\frac{M_p}{M_s}\right)^\frac{3}{2}
\left(\frac{a_B}{R_p}\right)^\frac{9}{2} 
\left( \frac{\mu_p Q_p}{GM_p^2/R_p^4} \right) \sqrt{\frac{M_p + M_s}{M_s}} 
\label{eqn:t_despinp}
\end{align}
where subscript $p$ refers to quantities for the primary, subscript $s$ 
refers to quantities for the secondary and subscript $B$ refers to quantities describing the binary mutual orbit.
In the case of a spinning secondary  
\begin{align}
\frac{t_{\rm despin,s}}{P_B} &\approx  
\left(\frac{M_s}{M_p}\right)^\frac{3}{2}
\left(\frac{a_B}{R_s}\right)^\frac{9}{2}
\left( \frac{\mu_s Q_s}{GM_s^2/R_s^4}\right)
\sqrt{\frac{M_p + M_s}{M_p}}.
\label{eqn:t_despins}
\end{align}

Wobble damping timescales estimated from asteroid light curves give an estimate for asteroid viscoelastic material properties $\mu Q \sim 10^{11}$ Pa \citep{harris94,Pravec_2014}.
Using this value for primary and secondary we estimate
\begin{align}
t_{\rm despin,p} &\approx 2.4\times 10^{11} \ {\rm  yr} 
\left(\frac{\mu_p Q_p}{10^{11}\ {\rm {Pa}}} \right) 
\left(\frac{a_B}{1.19 \ \mathrm{km}} \right)^3
\label{eqn:td1} \nonumber \\
& \times \left( \frac{q}{0.0093}\right)^{-2} \left( \frac{R_p}{390 \ {\rm m}} \right)^{-5}
\left( \frac{\rho_p}{2.1\ {\rm g\ cm}^{-3}} \right)^{- \frac{3}{2}}
\\
t_{\rm despin,s} &\approx 5.8 \times 10^{8} \ {\rm  yr}  \left(\frac{\mu_s Q_s}{10^{11}\ {\rm {Pa}}} \right) \left(\frac{a_B}{1.19 \ \mathrm{km}} \right)^3 \nonumber \\
& \times  \left( \frac{q}{0.0093}\right)^{2}
\left( \frac{R_s}{82 \ {\rm m}} \right)^{-5}
\left( \frac{\rho_s}{2.1\ {\rm g\ cm}^{-3}} \right)^{- \frac{3}{2}}
\label{eqn:td2}
\end{align}
using quantities measured for Didymos and Dimorphos listed in Table \ref{tab:Didymos}.
Here $q = M_s/M_p$ is the secondary to primary mass ratio. 


The long primary spin-down time in 
Equation \ref{eqn:td1} 
implies that tidal dissipation does not affect the primary's spin. 
All observed near-Earth and small main-belt asteroid binaries with secondary to primary diameter ratio lower than 0.6 have quickly spinning primaries 
with periods of a few hours  \citep{Pravec_2007,Pravec_2016}.  
On an asteroid, the reflection, absorption and emission of solar radiative energy produces a torque that changes the rotation rate and obliquity of a small body. 
The  Yarkovsky-O'Keefe-Radzievskii-Paddack (YORP) effect can account for the primary high spin rates \citep{Rubincam_2000,Scheeres_2015,Hirabayashi_2015}.    YORP could also slow down the primary's spin, though
at slow spin rates, 
the binary system may become unstable and disrupt, leading to fewer binaries with slowly spinning primaries.
For an illustration of an instability causing binary disruption, see Figure 14 by \citet{Davis_2020}.

The timescale in equation \ref{eqn:td2} for spin down time of the secondary is longer than the estimated NEA lifetime and is in conflict with the large number of binary asteroids with synchronous secondaries unless the binary asteroids were born in the main belt before they became NEAs or the tidal drift rate is underestimated by equation \ref{eqn:td2} \citep{taylor11}, as we discussed in section \ref{sec:intro}.

Constraints on asteroid internal properties based on tidal evolution of the orbital semi-major axis are only relevant if the BYORP induced drift rate is low (e.g., \citealt{taylor11}) or if the tidal dissipation is sufficiently strong that the tidal torque can counter the BYORP torque \citep{Jacobson_2011}.  The median value for small main belt binaries (primary radius $\sim 3$ km),  estimated with a collisional lifetime of $\Delta t = 10^9$ yr, 
is $\mu_p Q_p \sim 10^{11}$ Pa whereas that for small NEA binaries (with primary radius less than 1 km), 
estimated with a lifetime of $\Delta t = 10$ Myr is $\mu_p Q_p \sim 5 \times 10^8$ Pa \citep{taylor11}.  Assuming an equilibrium state between BYORP and tidal torque, 
\citet{Jacobson_2011} estimate $B Q_p/k_{2p} \sim 10^3$ 
giving $Q_p/k_{2p} \sim 10^6$. With BYORP coefficient $B\sim 10^{-3}$ this gives $Q_p \mu_p \sim 10^8$ Pa, similar to that estimated for NEA asteroid binaries by \citet{taylor11}.  
(Here we used equation \ref{eqn:k2} for the Love number, and estimated $0.038 G \left(4\pi/3 \rho_p R_p\right)^2 \sim 10^2$ Pa with values $R_p = 1$ km and $\rho_p =  2$ g/cm$^3$.)

As emphasized by \citet{wisdom87}, a non-round body must tumble before entering the spin-synchronous state.
Numerical simulations show that the time for obliquity and libration and non-principal axis rotation to damp can substantially prolong the time it takes to reach a spin synchronous state with low obliquity, low free libration and undergoing principal axis rotation \citep{Naidu_2015,quillen20_phobos}.
Hence the long spin down time estimated for Dimorphos in equation \ref{eqn:td2} suggests that Dimorphos could have spent a long period of time in a complex spin state that is nearly synchronous but has non-principal axis (NPA) rotation, obliquity oscillations and high amplitude libration.

Epochs of chaotic tumbling could prolong the lifetime of a binary asteroid as BYORP induced drift in the tumbling state is expected to be negligible.     
\citet{Naidu_2015} pointed out that the libration amplitude can be difficult to detect from a light curve.
With libration angle equal to the angle between the secondary's long axis and the vector
between primary and secondary centers of mass, 
   \citet{Pravec_2016} showed that libration angles greater than about $20^\circ$ can be detected through analysis of light-curve measurements.  \citet{Pravec_2016} attributed small offsets between the secondary rotational light curve and models to irregular secondary shape. While secondary light curves are sensitive to the orientation of the secondary's long axis, they may be less sensitive to 
 the orientation of the secondary's short axis.

\begin{figure} \centering
\includegraphics[width=3.2in,trim = 0 0 0 0, clip]{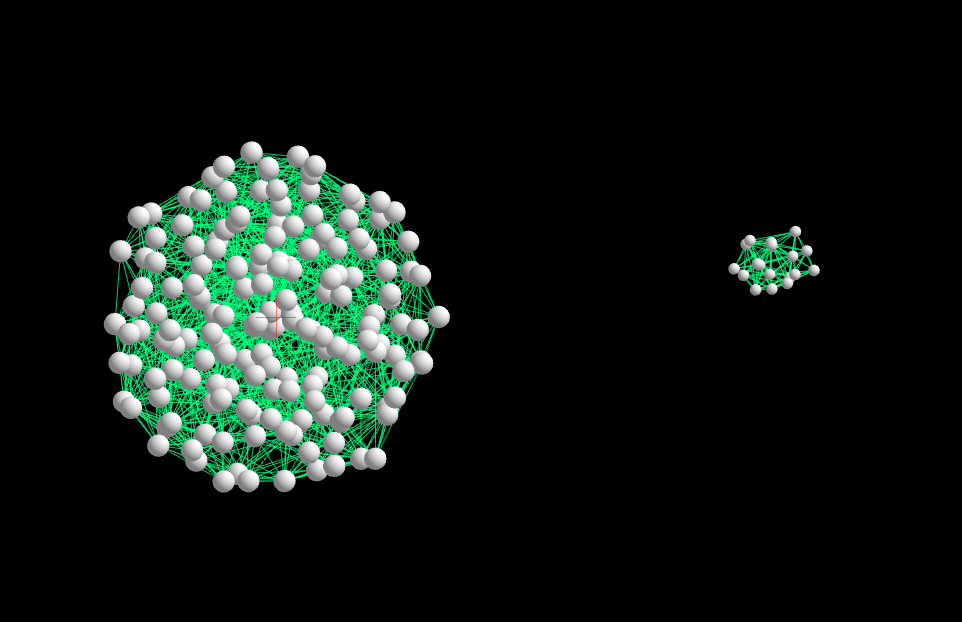}
\caption{A snap-shot of the mass/spring model simulation of a binary asteroid
 with output shown in Figure \ref{fig:b1}. 
The green lines show  springs and the gray spheres show mass nodes. Both primary  and secondary shapes
are resolved with mass nodes. \label{fig:snap}}
\end{figure}

\section{Mass/spring model simulations of an asteroid binary}
\label{sec:sims}

To explore the spin evolution of the two non-spherical bodies in an asteroid binary, we use the mass-spring model simulation code  \citep{quillen16_crust,frouard16,quillen16_haumea,quillen17_pluto,quillen19_moon,quillen19_wobble,quillen20_phobos},
that is built on the modular N-body code \texttt{rebound} \citep{rebound}.
In our code,  a viscoelastic solid is approximated as a collection of mass nodes that are connected by a network of springs.
Springs between mass nodes are damped and the spring network approximates the behavior of a Kelvin-Voigt viscoelastic solid. 

In our previous work we only resolved a single spinning body with masses and springs, though the body could be perturbed by point masses (e.g., \citealt{quillen17_pluto}).  Here we resolve both primary and secondary in the asteroid binary with masses and springs.   We do not include the effect of planets or tidal perturbations associated with the heliocentric orbit.  As we are interested in the behavior of long integrations (as when we studied the obliquity of Pluto and Charon's satellites; \citealt{quillen17_pluto}) rather than details of internal tidal dissipation \citep{quillen19_moon} we only use a dozen mass nodes to resolve the secondary and about a hundred mass nodes to resolve the primary.
The bodies are not spherical.  Because we resolve the bodies with individual mass nodes,  
we do not need to expand their gravitational perturbations in multipole terms.  See Figure \ref{fig:snap} for a snap-shot of one of our binary simulations.

The mass particles or nodes in both resolved spinning bodies are subjected to three types of forces:
the gravitational forces acting on every pair of mass nodes in the simulation, and the elastic and damping spring forces acting only between sufficiently close particle pairs.
As all forces are applied equal and oppositely to pairs of point particles and along the direction
of the vector connecting the pair, momentum and angular momentum conservation are assured.


We work with mass in units of $M_p$,  the mass of the primary in the asteroid binary,
distances in units of volumetric radius, $R_{\rm vol,p}$, the radius of a spherical body with the same
volume as the primary, and time in units of $t_{\rm grav}$ 
\begin{align}
t_{\rm grav} &\equiv \sqrt{\frac{R_{\rm vol,p}^3}{GM_p}} = \sqrt{ \frac{3}{4 \pi G \rho_p}} .
\label{eqn:tgrav}
\end{align}
This timescale at the density of the primary for the Didymos system is listed in Table \ref{tab:estimated}.

In our simulations, 
the primary and secondary are approximately the same density.  Spin is given in units
of $\omega_{\rm breakup} = t_{\rm grav}^{-1}$ (see equation \ref{eqn:om_breakup}) and computed using the mean density of the primary which is $3/(4 \pi)$ in our numerical units.  This frequency is also listed in Table \ref{tab:estimated} for the estimated density of Didymos.

Initial node distribution and spring network for each body is chosen with the tri-axial ellipsoid random spring model described by \citet{frouard16,quillen16_haumea}.
The confining surface used to generate the particles obeys $\frac{x^2}{a^2} + \frac{y^2}{b^2} + \frac{z^2}{c^2} = 1$
with $a,b,c$ equal to half the lengths of the desired principal body axes.  
Particles cannot be closer than a minimum distance $d_I$ from each other.
Since we generate the resolved bodies with only a few particles, axis ratios are more accurately computed
using the eigenvalues of the moment of inertia matrix after we generate the particle distribution.  For principal moments of inertia $A \ge B \ge C$, we compute the body axis ratios as 
$\frac{b}{a} = \sqrt{\frac{ A + C - B}{A + B -C}}$
and 
$\frac{c}{a} = \sqrt{\frac{B + C - A}{A+ B - C}}$.

Once the particle positions have been generated, every pair of particles within distance $d_{Spr}$ of each
other are connected with a single spring.  The parameter $d_{Spr}$ is the maximum rest length of any spring in the network.  
Springs have a spring constant $k_{Spr}$, giving elastic behavior and a damping rate parameter $\gamma_{Spr}$
giving viscous behavior \citep{frouard16}.
Our simulated material is compressible, so energy damping arises from both deviatoric and volumetric stresses.  
Springs are created at the beginning of the simulation and do not grow or fail during the simulation, so there is no plastic deformation.  
For numerical stability,
the simulation time-step must be chosen so that it is shorter than the time it takes elastic waves to travel between nodes \citep{frouard16,quillen16_haumea}.
All mass nodes in the primary have the same mass and all springs in the primary have the same spring constant and damping parameter. All mass nodes in the secondary have the same mass and all springs in the secondary have the same spring constant.  However, the mass nodes in the primary are not the same mass as those in the secondary.  Similarly the spring constants of the springs in the primary differ from those in the secondary.

\subsection{Simulation output}
\label{sec:simout}

At the beginning of the simulation we store the positions of the mass nodes for both primary and secondary.
The positions are in a coordinate system with $x$ axis aligned with the long principal body axis and $z$ axis aligned with the short principal body axis. The position of node $j$ from the center of mass of the body in which it resides is 
${\bf r}^0_j =(r^0_{j,0}, r^0_{j,1},r^0_{j,2})$
where $r^0_{j,0}$ is the $x$ coordinate.  
The position of node $j$ from the center of mass of the body in which it resides at a later time during the simulation is
${\bf r}_j =(r_{j,0}, r_{j,1},r_{j,2})$.
At each simulation output we compute the covariance matrix $S_{\alpha \beta} = \sum_j r^0_{j,\alpha} r_{j,\beta}$
using the current node positions  
and the initial node positions. The covariance matrix is stored so that we can later compute a rotation describing the body orientation at each simulation output using the Kabsch algorithm \citep{Kabsch_1976,Berthold_1986,Coutsias_2004}.  At each simulation output
the current body orientation is given 
by a rotation that we describe with a unit quaternion.  The quaternion rotates the initial body node positions
to give the node positions  at the simulation output time.
This quaternion is equivalent to the unit eigenvector associated with the maximum eigenvalue of a 
4x4  matrix derived from the covariance matrix $S_{\alpha \beta}$ \citep{Berthold_1986,Coutsias_2004}.  
From the covariance matrix at each simulation output  we compute a quaternion at each simulation output time.  We use this quaternion to compute the orientation of the body's long and short axes by rotating the $x$ and $z$ axes at each simulation output time.

At each simulation output time we compute the primary and secondary's moment of inertia matrix from the positions  of their mass nodes with respect to the body's center of mass.   
The spin angular momenta for each body, ${\bf L}_s, {\bf L}_p$, are computed from the positions 
and velocities of each body's mass nodes.   
The spin vectors for each body, ${\boldsymbol \omega}_s, {\boldsymbol \omega}_p$
 are computed from the body spin angular momentum
vectors and the inverse of their moment of inertia matrices, e.g., 
${\boldsymbol \omega}_s = {\bf I}_s^{-1}{\bf L}_s$.

We denote the directions of the secondary's long and short 
body principal axes as $\hat {\bf i}_s$ and $ \hat {\bf k}_s$, respectively.  The
intermediate secondary body principal axis is $\hat {\bf j}_s$.    
The inclination of the secondary's long principal axis is the angle between $\hat {\bf i}_s$
and the binary orbital plane.    The secondary's non-principal angle $\theta_{\rm NPA,s}$ is
the angle between the secondary's short principal axis $ \hat {\bf k}_s$
and the direction of the secondary's spin angular momentum, ${\bf L}_s$. 
The secondary's short axis tilt is the angle
between the secondary's short axis and the orbit normal, 
${\rm arccos} (\hat {\bf k}_s \cdot \hat {\bf l}_B)$.    The orbit normal is close to the $z$ axis during the simulation.
The secondary's precession angle $\theta_{ls}$ is the angle on the 
$xy$ plane of the secondary's spin angular momentum vector, $\theta_{ls} = {\rm arctan2}(L_{sy}, L_{sx}$)
where ${\bf L}_s  = (L_{sx},L_{sy}, L_{sz})$.
The obliquities for primary and secondary $\epsilon_s, \epsilon_p$ are the angle between body spin angular momentum and orbital normal $\hat {\bf l}_B$.  
The secondary libration angle $\phi_{lib,s}$ is that between the secondary's long principal body axis, projected onto the orbital plane, and the vector connecting primary and secondary centers of mass.

The binary orbit normal $\hat{\bf l}_B$ direction is computed using the body center of mass  positions and velocities.    
The orbital semi-major axis $a_B$, eccentricity $e_B$ and inclination $i_B$ are computed from the positions and velocities of the primary and secondary center of mass and assuming Keplerian motion (we assume point masses so the extended mass distributions are not taken into account).
The orbital inclination is zero if the binary orbit lies in the $xy$ plane.
We measure the angle of the secondary in its orbit, $\theta_B$,  from the vector between the two body center of mass positions, $\bf r$,  projected into the $xy$ plane.  As orbital inclination and eccentricity
are low, $\theta_B = {\rm arctan2}(r_y, r_x) $ is approximately equal to the binary's mean longitude, $\lambda_B$.  
Using an epicyclic approximation we measure an epicylic angle $\theta_{kB}$ describing
radial oscillations of the orbit.  
The epicyclic angle is computed from the vector between secondary and primary,  the radius $r = |{\bf r}|$, 
the mean value of the radius, $\bar r$,   the orbital mean motion $n_B$, 
and the radial velocity component,  $v_r = \frac{dr}{dt}$.  
The mean radius $\bar r$ and mean motion are estimated by smoothing the time series
of $r$ and $n_B$ with a Savinsky-Golay filter that has a window length of a few orbits.
The epicyclic angle is then $\theta_{kB} = {\rm arctan2}((r - \bar r)n_B, v_r)$.
The angles $\theta_B, \theta_{kB}$ are used instead of orbital elements 
because a Keplerian approximation is poor due to the non-spherical shapes  of the primary and secondary (e.g., \citealt{Renner_2006,Agrusa_2020}).

\subsubsection{Multipole coefficients}

We describe our convention for calculating gravitational multipole coefficients for the primary and secondary.
Following \citet{kaula66,Tricarico_2021},
the gravitational potential outside a massive non-spherical body,  of mass $M$, where it satisfies Laplace's equation, can be expanded in terms of multipoles, 
\begin{align}
V(r,\theta,\phi) & = -\frac{GM}{r} \Bigg[ 1 + \sum_{l=2}^\infty
\sum_{m=0}^l  \left( \frac{R_{vol}}{r} \right)^{l}
 P_{lm} ( \cos \theta)  \times \nonumber \\
& \ \ \ \ \ \ \ \ \ \ \ \ \ \ \  \left(  C_{lm} \cos (m \phi)  + S_{lm} \sin (m \phi) \right)  \label{eqn:V}
\Bigg]
\end{align}
where $r$ is radius,  $\phi \in [0, 2\pi]$ is longitude and $\theta \in [0, \pi]$ is colatitude.
The center of mass is at the origin and this is why the $l=1$ terms are absent from the expansion.
The functions $P_{lm}()$ are the associated Legendre polynomials
which can be computed using Rodrigues' formula
\begin{equation}
P_{lm}(x) = \frac{(1-x^2)^{m/2}}{2^l l!} \frac{d^{l+m}}{dx^{l+m}} (x^2 - 1)^l.
\end{equation}
The dimensionless multipole coefficients, $C_{lm}, S_{lm}$,  depend on a characteristic   
 length  which we set to be that of the volume equivalent sphere $R_{vol}$.
Each mass node in our simulations is specified with an index $i$,  has mass $m_i$ and spherical coordinates  $(r_i,\theta_i, \phi_i)$.
The multipole coefficients of the body are
\begin{align}
 C_{lm} & = \frac{1}{M}  \frac{(l-m)!}{(l+m)!} (2 - \delta_{m0})   \sum_i m_i \left(\frac{r_i}{R_{vol}} \right)^l  
 P_{lm} (\cos \theta_i) \cos (m \phi_i)  \nonumber  \\
 S_{lm} & = \frac{1}{M} \frac{(l-m)!}{(l+m)!} 2   \sum_i  m_i \left(\frac{r_i}{R_{vol}} \right)^l  
  P_{lm} (\cos \theta_i) \sin(m \phi_i) 
 \label{eqn:cofs}
\end{align}
where total mass $M = \sum_i m_i$.
The $l=0$ coefficients are called zonal coefficients, $J_l = -C_{l0}$,  
and the $l\ne0$ coefficients are called tesseral coefficients.

\begin{table*}[]
    \centering
    \caption{\Large Viscoelastic Mass/Spring Model Simulation Parameters for Tidal spin down}
    \label{tab:sim}
\begin{tabular}{llll}
\hline
Description & Symbol & Quantity \\
\hline
 Total integration time &   $t_{max}$    & $5 \times 10^5 = 15400\ P_B$ \\
 Time step &   $dt$ & 0.002 \\
 Initial orbital semi-major axis & $a_B$ & 3.0 \\
 \hline
 Description & Symbols & For Primary & For Secondary \\
 \hline
 Mass    & $M_p, M_s$ & 1 & 0.01 \\
 Radius equiv. volume sphere & $R_{vol}$ & 1 & 0.22 \\
 Initial spin & $\omega_{init}$ &  0.8 & 0.5  \\
 Minimum interparticle distances & 
 $d_I$ & 0.27 & 0.14\\
 Maximum spring length/$d_I$ & $d_{Spr}/d_I$ & 2.4 & 2.3\\
 Spring constants & $k_{Spr}$ & 0.20 & 0.03\\
 Spring damping parameters & $\gamma_{Spr}$ & 1 & 30 \\
Body axis ratios  & $a:b:c$ &1.0:0.958:0.881 & 1.0:0.773:0.698\\
 Number of particles & $N$ & 165 & 18 \\
 Number of springs  & $N_{Spr}$ & 1950 & 113 \\
\hline
\end{tabular} \\
Axis ratios are those derived from the principal axis moments of inertia and that are computed from the generated particle positions. Time is given in units of $t_{\rm grav}$.   Distances are in units of $R_p$, the primary's
volume equivalent radius.
Spin rates are in units of the break up spin rate  $\omega_{\rm breakup}$ or  equivalently $t_{\rm grav}^{-1}$.  Multipole coefficients for primary and secondary are listed in Table \ref{tab:multipole}.
\end{table*}

\subsection{Simulation of secondary tidal spin down}
\label{sec:sim1}

We carry out a simulation  with simulation parameters listed in Table \ref{tab:sim}. 
The secondary is initially spinning quickly and is not in a spin-orbit resonance.
Secondary to primary mass ratio, body axis ratios, primary spin, and binary orbital semi-major axis are intended to be similar to those of  the Didymos system.
 The numbers of mass nodes and springs in each body and 
ellipsoid axis ratios measured from the body's moments of inertia are  
listed in Table \ref{tab:sim}.
For this simulation we set the damping parameters for the springs to be at relatively high values, 
following \citet{quillen17_pluto,quillen20_phobos}, so that we can see the process of tidal spin down due to dissipation.  

Both bodies were initially set to be rotating
about their shortest principal axis and at zero obliquity. 
Here obliquity is the angle between body spin angular momentum and binary orbit normal vector direction $\hat {\bf l}_B$.
The binary orbit is initially in the $xy$ plane with orbit normal $\hat {\bf l}_B = \hat {\bf z}$  in the $z$ direction. 
We set the spin of the primary to be $\omega_p = 0.8\ \omega_{\rm breakup} = 0.8\ t_{\rm grav}^{-1}$, 
below the near 1 value that Didymos currently has, 
to ensure that self-gravity can hold the primary together while keeping the springs under slight compression. 
As springs do not dissolve in our simulation, a rapidly spinning body near rotational breakup would unphysicaly put the springs under tension.  The secondary is initially at $\omega_s = 0.5\ \omega_{\rm breakup}$.

The binary orbit was begun at low orbital eccentricity.  
The initial orbit eccentricity $e_B$ is not exactly zero because
we compute it without taking into account the non-spherical mass distributions of primary or secondary
(for discussion on this issue see  \citealt{Agrusa_2020}).

\begin{figure} \centering
\includegraphics[width=3.5in, trim=30 30 10 0, clip]{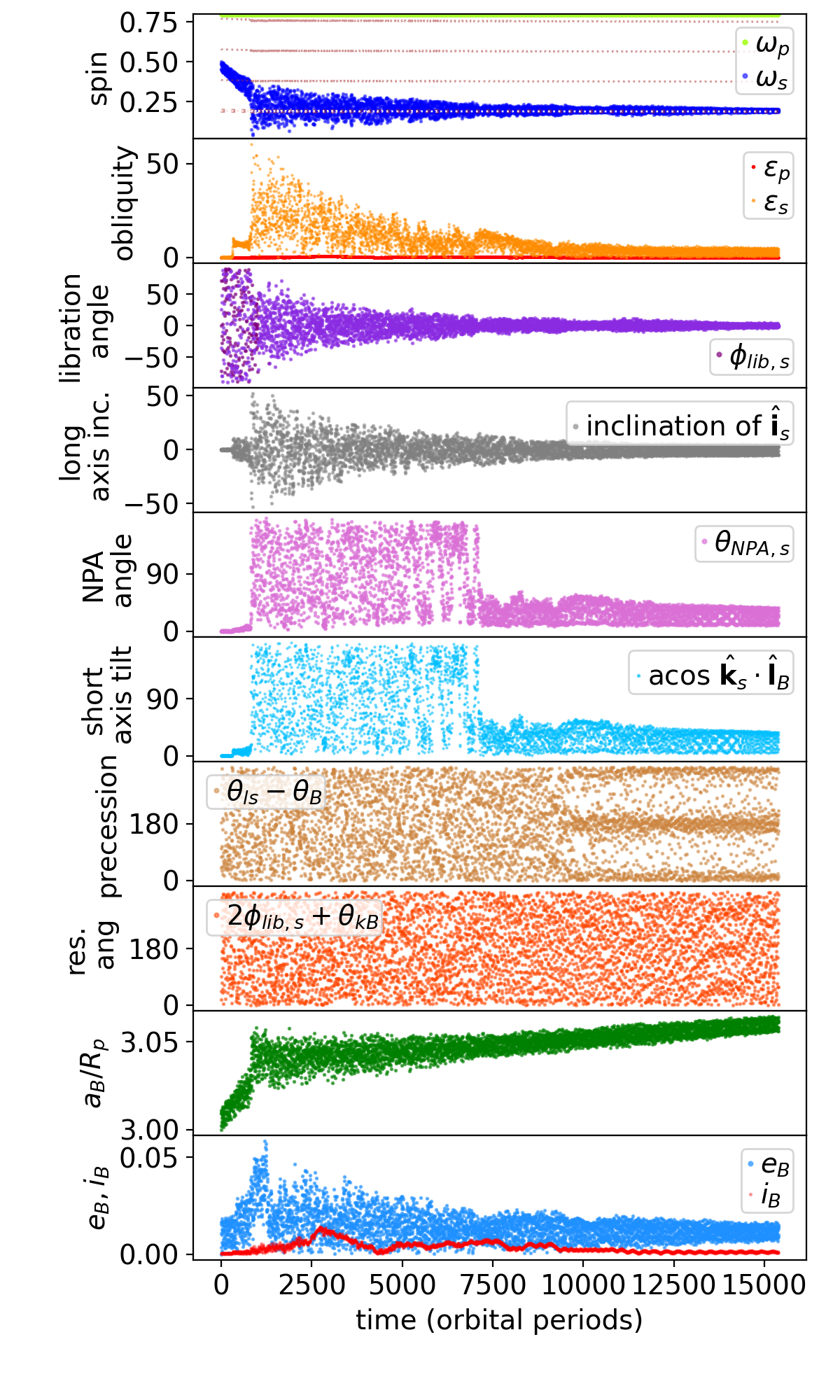}
\caption{Simulation of a tidally evolving asteroid binary.  Both primary and secondary are
extended non-spherical bodies.
The parameters for this simulation are listed in Table \ref{tab:sim} and a simulation snap-shot is shown
in Figure \ref{fig:snap}.
The top panel shows primary and secondary spin rates in units of $t_{\rm grav}^{-1}$.  
The second panel from the top shows primary and secondary obliquity 
(which is the angle between spin angular momentum and orbit normal).   
The third panel from top shows the secondary libration angle,  $\phi_{lib,s}$.  
In the middle panels, 
we  show the inclination of the secondary's long body axis with respect to the orbital plane, 
the secondary's non-principal axis angle $\theta_{\rm NPA,s}$, the tilt of the secondary's short axis with
respect to the orbit normal ${\rm acos} (\hat{\bf k} \cdot \hat {\bf l}_B) $,
 a resonant angle associated with precession and a resonant angle associated with libration. 
The second from bottom panel  shows the binary orbit semi-major axis 
in units of the primary's radius, $a_B/R_p$. 
The bottom panel shows binary orbital eccentricity $e_B$ and inclination $i_B$.
Angles are in degrees except for the binary orbit inclination in the bottom panel which is 
in radians. 
Time is in units of binary's initial orbital period.
This simulation illustrates that the spin-down rate of the secondary as it approaches the spin synchronous state (shown in the top panel at early times) is faster than the obliquity and non-principal axis 
rotation damping rate at later times.
\label{fig:b1}
}
\end{figure}

In Figure \ref{fig:b1}, we plot primary and secondary spin in units of $t_{grav}^{-1}$,  obliquity, secondary libration angle, orbital semi-major axis, eccentricity and inclination as a function of time. 
The horizontal axis shows time in units of the initial binary orbital period, $P_B$.
All angles are in degrees except for the orbital inclination $i_B$ which is in radians.
In this figure we also plot the secondary's long axis inclination angle,
non-principal axis angle $\theta_{\rm NPA,s}$, short axis tilt angle, and resonant angles
associated with precession and libration.


In the top panel of Figure \ref{fig:b1}, we plot both primary and secondary spins.  
Dotted lines show the locations of spin-orbit resonances where the spin is equal to the binary orbit mean motion $\omega = n_B/j $ divided by integer $j$.
The secondary enters the 1:1 spin-orbit resonance at about $t\approx 1000\ P_B$ at which time 
the libration angle $\phi_{lib,s}$ begins to librate about 0.  Simultaneously, the secondary's obliquity
jumps to about $\epsilon_s \sim 40^\circ$ and non-principal
axis rotation is excited.  The non-principal axis rotation
  is reflected in the large range exhibited by the short axis tilt angle.
A smaller jump in obliquity occurs at about $t \sim 400\ P_B$ where
the 2:1 spin-orbit resonance (where $\omega_s \approx 2 n_B$) is crossed.
Because tidal dissipation also is present in the primary, the primary spin 
slowly decreases during the simulation and this is accompanied by a slow increase in the orbital semi-major axis.

\subsubsection{Non-principal axis rotation and the time it takes for it to decay}

The simulation shown in Figure \ref{fig:b1} illustrates that the spin-down rate of the secondary (shown in the top panel) is faster than the obliquity and non-principal axis damping rate. Past a time of about 1000 orbit periods,  
the secondary spin rate on average is consistent with being in the 1:1 spin-orbit resonance,   with secondary spin rate equal to the binary orbit mean motion $\omega_s \sim n_B$.  Long after ($10^4$ orbit periods after)  the secondary enters the 1:1 spin-orbit resonance, its obliquity and non-principal axis rotation angle $\theta_{\rm NPA,s}$ remain high and vary chaotically.  Long lasting obliquity and non-principal axis angle variations were seen in some simulations of tidally evolving Phobos and Deimos \citep{quillen20_phobos}, though not with as high values of $\theta_{NPA}$.  We confirm the findings of 
\citet{Naidu_2015} who similarly emphasized that chaotic non-principal axis rotation of the secondary could be long-lived in binary asteroid systems, though they discussed tumbling rather than NPA rotation within the 1:1 spin orbit resonance.

\begin{figure} \centering
\includegraphics[width=3.5in, trim=0 0 0 0, clip]{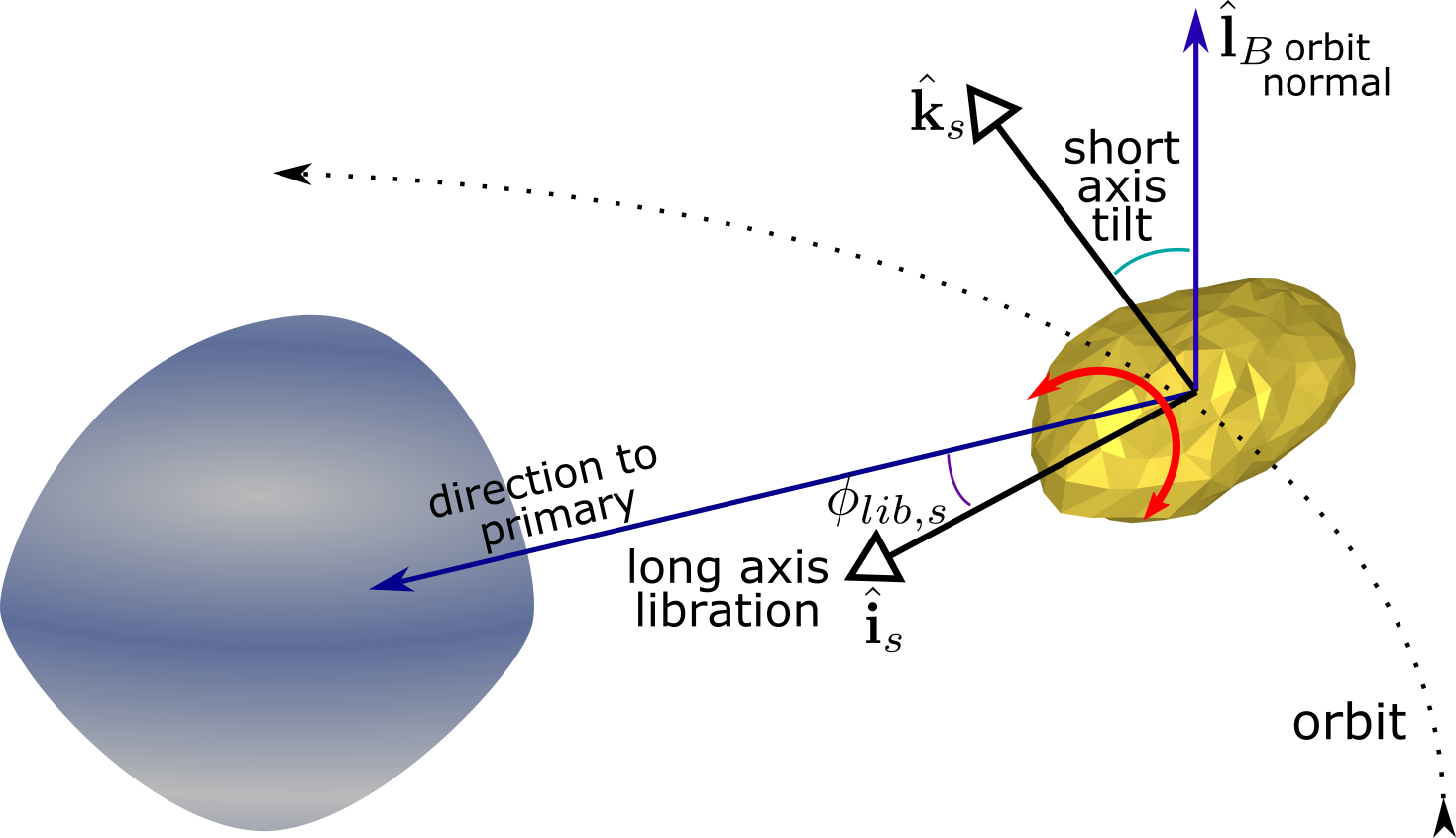}
\caption{An illustration of the non-principal axis rotation state of the secondary for the simulation
shown in Figure \ref{fig:b1} after the secondary has entered the 1:1 spin-orbit resonance.  
The secondary's long axis (shown
with direction $\hat {\bf i}_s$) remains aligned along 
the direction to the primary and this maintains a low libration angle $\phi_{lib,s}$.   
However, the secondary's short axis, with direction $\hat {\bf k}_s$, 
can rotate as shown with red arrow.  The secondary can slowly rotate about its long axis 
without leaving the 1:1 spin-orbit resonance.    
The short axis tilt angle is the angle between  $\hat {\bf k}_s$ and  
the binary orbit normal  $\hat {\bf l}_B$.  The short axis can rock back and forth, as shown by
the red arrow,  due to non-principal axis rotation.  
Replacing the secondary with an airplane that has 
nose pointing to the primary, long axis libration is equivalent to {\it yaw}, short axis tilt is equivalent to {\it roll}
and variations in the long axis inclination are equivalent to {\it pitch}.
\label{fig:illust}
}
\end{figure}

The nature of the secondary's non-principal axis rotation state in the simulation 
shown in Figure \ref{fig:b1} is illustrated in Figure \ref{fig:illust}.  The secondary's long body axis 
remains aligned with the direction to the primary giving it a low libration angle that oscillates
about zero.   Meanwhile the short axis of the secondary can rotate. The body can rotate about its long
axis without significantly increasing the libration angle.   
The secondary lies in the 1:1 spin-orbit resonance 
 but also exhibits significant non-principal axis rotation.  The light curve of such a secondary might appear
to reside in a low obliquity and tidally locked principal axis rotation state.  It might be difficult to 
differentiate between a principal axis rotation state and  a non-principal axis rotation state if the libration angle 
$\phi_{lib,s}$ remains low in both states.

In a frame synchronous with the orbit and replacing the secondary with a plane that has 
nose pointing to the primary, long axis libration is equivalent to {\it yaw}, short axis tilt is equivalent to {\it roll} and variations in the long axis inclination are equivalent to {\it pitch}.
Independently and concurrent with our study,  
\citet{Agrusa_2021} have also noticed that within the 1:1 spin-orbit resonance, an elongated secondary 
can rotate about its long axis.   This may be related to a phenomenon called the ``Barrel instability"   \citep{Cuk_2020}. 
 
Tidal spin down in our simulation takes place in 1000 orbital periods, corresponding to only 1.4 years
at the binary orbital period of Didymos.  This is about 8 orders of magnitude shorter than
estimated for Dimorphos' spin down via equation \ref{eqn:t_despins}.
The ratio of the tidal spin down time to the binary orbital period of 1000 is long enough that the system
 could be varying adiabatically. However, the tidal dissipation rate is fast enough that the simulation may not exhibit phenomena associated with weaker resonances that only would be noticed
at lower drift rates \citep{quillen06_rescap}.  
Below we compare the non-principal axis rotation damping time to that of spin down.
Even if we are not well into the adiabatic regime, the two phenomena may
exhibit similar sensitivity to the tidal dissipation rate. 
Multiple simulations of the same body (with axis ratios similar to those of our secondary)
that were begun at slightly different initial conditions   
 showed a wider variety of phenomena in slowly drifting simulations than
 in more rapidly drifting ones  \citep{quillen20_phobos}.
Future work could check that longer more adiabatically drifting simulations show similar phenomena as we find here and could quantitatively explore the diversity of possible secondary NPA rotation exhibited within the 1:1 resonance.

The study of Phobos and Deimos, which compared tumbling decay 
at different tidal dissipation rates \citep{quillen20_phobos}, found that non-principal axis rotation
is sometimes prolonged at lower tidal drift rates, however, the ratio of 
time spent within the 1:1 spin orbit
resonance at high levels of NPA rotation to that required for spin down was typically much shorter than
we see in our binary asteroid simulation. 
The simulated Phobos and Deimos have similar body axis ratios as the secondary simulated here, so
body shape does not account for the difference in length of time spent at high levels of NPA rotation
within the 1:1 resonance.
The Phobos and Deimos simulations shown 
in Figures 4 and 5 by \citet{quillen20_phobos}
have a larger ratio of orbital semi-major axis to moon radius ($a/R_{moon} \sim 48$) than the
equivalent ratio for the binary asteroid simulations presented here  ($a_B/R_s \sim 14$, 
which is similar to the actual value).  These Phobos and Deimos simulations were done
 with Mars modeled by a point mass.  However, our simulated asteroid secondary orbits
a non-round primary with ratio of orbital major axis to primary radius $a_B/R_p \sim 3$ 
which is similar to the actual value for the Didydmos system.
The ratios $a_B/R_s$ and $a_B/R_p$ are 
low enough in binary asteroid systems
that the non-spherical body shapes could affect the spin dynamics \citep{Agrusa_2020}. 
Higher order multipole components in the gravitational potential are effectively stronger
in the binary asteroid than the Phobos/Deimos setting due to the non-spherical shapes of both primary and secondary and their proximity.  
Mars has zonal  coefficient $J_2 \approx 8.75 \times 10^{-4}$ \citep{genova16}, whereas 
asteroid (101955) Bennu has $J_2 \approx 0.01926$ \citep{Chesley_2020}.
The multipole coefficients of the primary in a binary asteroid system are likely to be more than an order 
larger than those of Mars.

\begin{figure} \centering
\includegraphics[width=3.5in, trim=30 30 10 0, clip]{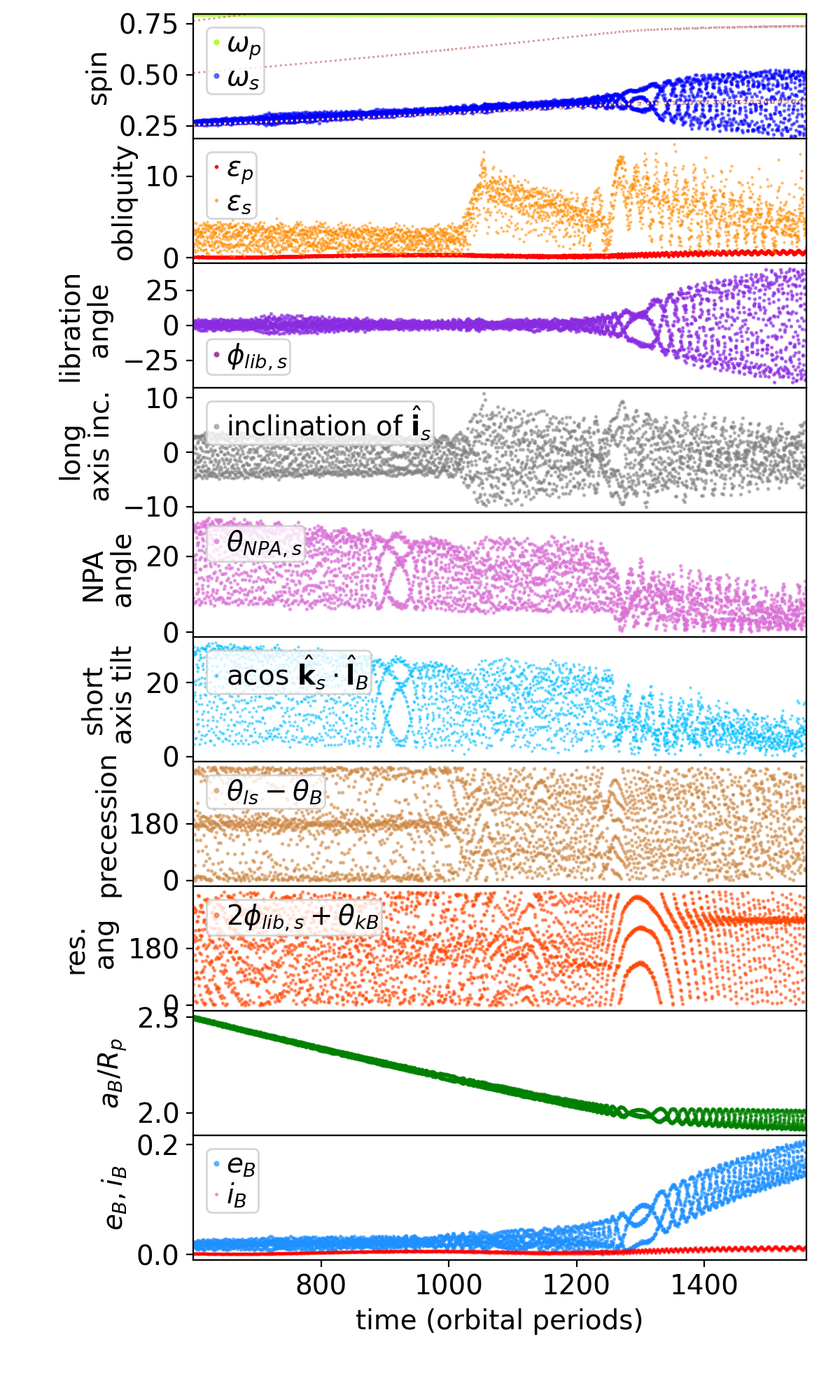}
\caption{A binary asteroid simulation that exhibits forced inward migration.
The initial conditions for this simulation are given by the last output of the simulation
shown in Figure \ref{fig:b1}.   The panels of this figure are similar to those of Figure \ref{fig:b1}.
Time is given in orbital periods from the beginning of the simulation.  
This simulation illustrates that secondary obliquity and non-principal axis rotation can be excited during inward
orbital migration.  
At $t\sim 1070\ P_B$ the obliquity is excited, the long axis is tilted out of the orbital
plane  and  the resonant
angle $\theta_{ls} - \theta_B$  ceases spending time at 0 and $180^\circ$. At $t\sim 1300\ P_B$ 
a libration resonance is excited with resonant angle $2\phi_{lib} + \theta_{kB}$.
The orbital eccentricity increases and the binary disrupts at $t \sim 1600\ P_B$.
This figure shows that obliquity and non-principal axis rotation 
can be excited in the secondary via induced migration.
\label{fig:b1c}}
\end{figure}

Because the multipole terms in the gravitational field may be important we have
computed them using equations \ref{eqn:cofs} for primary and secondary in
the simulation shown in Figures \ref{fig:b1} and \ref{fig:b1c} and list them for $l=2$ to 5
in Table \ref{tab:multipole}.
The coefficients are computed with origin at the body's center of mass,  with $x$-axis oriented along the longest body principal axis and $z$-axis oriented along the shortest principal axis. 
This orientation gives $C_{21} = S_{21} \approx 0$.  Zonal coefficients $J_{l} = - C_{l0}$.

\begin{table*}
\caption{\Large Multipole coefficients of primary and secondary \label{tab:multipole}}
\begin{adjustbox}{width=6.5in,center}
\begin{tabular}{llllll}
\hline
primary \\
\hline
$C_{20}=$-4.6333E-02 & $C_{21}=$1.2341E-08 & $C_{22}=$5.1758E-03 \\ 
                                     & $S_{21}=$-1.3732E-08 & $S_{22}=$-1.2697E-08 \\
$C_{30}=$8.6079E-03 & $C_{31}=$2.0366E-03 & $C_{32}=$3.3490E-04 & $C_{33}=$4.1048E-04 \\ 
                                    & $S_{31}=$-3.3649E-03 & $S_{32}=$-4.1383E-04 & $S_{33}=$5.3149E-04\\ 
$C_{40}=$2.3331E-03 & $C_{41}=$-2.7991E-04 & $C_{42}=$-4.5127E-04 & $C_{43}=$-5.1480E-05 & $C_{44}=$-2.2386E-05 \\ 
                                    & $S_{41}=$-8.2225E-04 & $S_{42}=$-2.9561E-04 & $S_{43}=$1.2788E-04 & $S_{44}=$-1.1213E-05 \\ 
$C_{50}=$5.8134E-03 & $C_{51}=$-9.6107E-04 & $C_{52}=$-6.0360E-05 & $C_{53}=$5.0140E-05 & $C_{54}=$2.4168E-05 & $C_{55}=$6.6303E-06 \\ 
                                    & $S_{51}=$1.3849e-03 & $S_{52}=$4.9828E-06 & $S_{53}=$-4.1337E-05 & $S_{54}=$-1.4656E-05 & $S_{55}=$3.8767E-06 \\                                 
\hline
secondary \\
\hline
$C_{20}=$-1.2240E-01 & $C_{21}=$-1.8134E-07 & $C_{22}=$3.9505E-02 \\ 
                                     & $S_{21}=$-1.1728E-07 & $S_{22}=$9.9006E-08 \\ 
$C_{30}=$1.6585E-02 & $C_{31}=$1.6529E-02 & $C_{32}=$6.9445E-03 & $C_{33}=$9.8851E-05 \\ 
                                    & $S_{31}=$6.5180E-03 & $S_{32}=$2.2701E-03 & $S_{33}=$-1.1486E-03 \\ 
$C_{40}=$2.1781E-02 & $C_{41}=$-9.5380E-03 & $C_{42}=$-3.3054E-03 & $C_{43}=$-1.6999E-03 & $C_{44}=$2.9416E-04 \\ 
                                     & $S_{41}=$-7.2263E-03 & $S_{42}=$-1.0916E-03 & $S_{43}=$1.7710E-03 & $S_{44}=$1.5515E-04\\ 
$C_{50}=$1.8595E-02 & $C_{51}=$-6.2283E-03 & $C_{52}=$-1.0997E-03 & $C_{53}=$6.9053E-04 & $C_{54}=$2.8126E-04 & $C_{55}=$2.8868E-05 \\ 
                                    & $S_{51}=$6.0582E-03 & $S_{52}=$2.0111E-04 & $S_{53}=$1.5493E-04 & $S_{54}=$7.4877E-05 & $S_{55}=$7.5874E-05 \\ 
\hline
\end{tabular}
\end{adjustbox}
\end{table*}

The primary in our simulations has multipole coefficients that have similar magnitude 
to those of asteroid Bennu (in Table 5 by \citealt{Chesley_2020}).  
The multipole coefficients of the secondary are larger than those of the primary because it is more elongated and because it is resolved in the simulation with fewer mass nodes.  
Are the multipole coefficients sufficiently strong that they could affect the spin dynamics of the secondary in our simulations?  
The primary is spinning quickly.  Perturbations on the secondary associated with the multipoles of the primary 
would have  frequencies that exceed the secondary's spin.  However, 
via a higher order expansion, perturbations with frequencies that are sums or differences
of other perturbation frequencies (as are commonly found in second order expansions) 
might be similar to the rotation rate and so might account
for the chaotic behavior.
We tentatively attribute the long lived non-principal axis rotation seen in our simulations 
to the strength of perturbations associated with multipole potential terms that are present  
 because both primary and secondary are non-spherical.   

Figure \ref{fig:b1} shows that 
the time to damp the non-principal axis rotation and obliquity in our simulation is longer than the time it takes
for the secondary to spin down and enter the spin synchronous state.   
Fitting time dependent exponentially decreasing functions to smoothed secondary spin and obliquity arrays, we find that the damping time is about 5 times longer for obliquity after reaching the 1:1 spin-orbit resonance than for spin prior to reaching the resonance. 

As discussed in section \ref{sec:spindown}, the time for the secondary to spin down could be similar  to the lifetime of NEA asteroids. 
With a value considered typical of main belt asteroids, the tidal spin down time for Dimorphos exceeds the NEA lifetime.
If binary asteroid secondaries are predominantly in principal axis rotation states, then much lower values of $\mu Q$ would be required to account for their synchronous rotation.   
This follows because it takes longer for tidal torque to reduce the level of NPA rotation after reaching the 1:1 spin orbit resonance. 
In our simulation, we have compared the decay
time of the non-principal axis rotation to the tidal spin down time.   Assuming that the tidal dissipation  
rate in the simulations is sufficiently adiabatic that the ratio
of the two time scales  is insensitive to the tidal dissipation rate, we infer that it takes even longer
for NPA rotation to decay than to enter the 1:1 spin orbit resonance. 
The even longer time require for non-principal axis rotation damping exacerbates the difficulty  reconciling the estimates for 
$\mu Q$ in binary and single asteroids.   

What if the asteroid binary secondaries are 
commonly within complex rotation states with non-principal axis rotation?  They could still reside
in the 1:1 spin-orbit resonance while exhibiting non-principal axis rotation, as seen in our simulation
shown in Figure \ref{fig:b1}.   If binary asteroid secondaries are commonly found in non-principal axis rotation states 
then their tidal dissipation rate cannot be extremely high.    Constraints on secondary rotation state
may improve upon tidal based estimates for their material properties. 

In Figure \ref{fig:b1}, in the panel labeled `precession' (fourth from bottom), we show
a resonant angle $\theta_{ls} - \theta_B$ which is the secondary spin angular momentum direction
projected onto the orbital plane as seen in the frame rotating with the secondary in its orbit.  
 At $t \sim 9500\ P_B$ this resonant angle is mostly near 
 zero and $180^\circ$ and at the same time there is a jump in the secondary's non-principal axis angle $\theta_{\rm NPA,s}$.  
 The distribution of $\theta_{ls} - \theta_B$ 
 suggests that the short axis tilts back and forth with an oscillation period that is
 twice the binary orbit period. 
 The secondary's spin angular momentum ${\bf L}_s$  is often  
 aligned with the direction between secondary and primary.  
At the same time the secondary's short axis, when projected
onto the $xy$-plane,  spends time at angle $\pm 90^\circ$. 
The secondary's short axis periodically rocks back and forth about the orbit normal because of
rotation about its long axis.
The jump in the non-principal axis angle $\theta_{\rm NPA,s}$
suggests that it is a resonant state, so it could last a really long time (10 to 100 times longer than the spin down time, as did 
resonant non-principal axis rotation states seen in simulations of Deimos;  \citealt{quillen20_phobos}).
In future,  the theoretical framework developed by \citet{Boue_2009,Fahnestock_2008,Agrusa_2021} for binary systems might be applied to better identify and create a model for this NPA rotation resonance.
If sub-resonances are identified within the 1:1 spin-orbit resonances, then perhaps a theory accounting
for the high level of NPA rotation exhibited by our simulations could be developed.

We explored simulations similar to that shown in Figure \ref{fig:b1}  at larger 
orbital semi-major axis with a larger secondary to primary mass ratio 
(similar to that of the binary system Binary Near-Earth Asteroid (66391) 1999 KW4 now known as Moshup) and saw similar behavior to that described here.  
We also explored similar simulations with  different axis ratios for the secondary (within 0.1
of those used here)  we and also saw similar behavior.

\subsection{Simulations of inward migration}
\label{sec:mig}

YORP and BYORP effect torques cause slow evolution so they can be considered adiabatic.   
BYORP effect torque could cause the binary orbital semi-major axis to drift 
either outward or inward \citep{Cuk_2005}.
We explore whether obliquity and non-principal axis rotation can be excited during orbital migration. 
We carry out a simulation that forces the binary orbital semi-major axis to shrink 
with a drift rate $\dot a_B = -10^{-5} R_p/t_g$. 
Migration is forced by perturbing the center of mass velocities 
using the procedure described by \citet{Beauge_2006} and which 
we used previously to force outward migration of Charon when studying the obliquity
evolution of Pluto and Charon's minor satellites \citep{quillen17_pluto}. 
Our migration simulation uses the last output of the 
tidal spin down simulation (shown in Figure \ref{fig:b1}) as its initial conditions.   Thus its parameters
 are the same as those listed in Table \ref{tab:sim} except that we set 
 the spring damping constants to 0.01 so as to reduce tidal dissipation. 
The secondary begins in the 1:1 spin synchronous state but with some NPA rotation.  
The migration simulation was run until the binary disrupted at $t \sim  1600\ P_B$.  
The simulation is shown in Figure \ref{fig:b1c} which has panels similar
to those of Figure \ref{fig:b1}.
We restricted the time shown in the Figure to make it easier to see
variations in angles and other quantities prior to the short period of instability that led to its breakup.

Figure \ref{fig:b1c} shows two obliquity jumps, one at $t \sim 1070\ P_B$ and the 
other at $ t\sim 1300\ P_B$.  The first jump in obliquity occurs when the precession angle
$\theta_{ls} - \theta_B$ ceases to be near zero
and $180^\circ$ and spends more time near $90^\circ$ and $270^\circ$.
At the same time the secondary's long axis is tilted out of the orbital plane giving oscillations in the 
long axis inclination angle.    The body's long axis tilts above and below the orbital plane
with an oscillation period that is equal to 2 binary orbit rotation periods. 
In a frame rotating with the secondary in its orbit, 
the secondary shifts from rotating about its long axis (and tilting its short axis w.r.t to the orbit normal)
to rotating about its intermediate axis and giving variations in the long axis inclination.  

The second jump in obliquity at $t \sim 1300 P_B$ is associated with an increase in the libration angle and orbital eccentricity.
This second jump is also associated with a transition in the behavior or the resonant angle $2 \phi_{lib,s} + \theta_{kB}$.
Prior to the second transition, this angle oscillates and afterwards it librates about $270^\circ$.
This angle corresponds to the secondary rotating twice in the frame of orbital rotation about its short axis per epicyclic period.
The resonance is strong enough to increase the orbital
eccentricity until the binary disrupts. 
We suspect the disruption mechanism is the same as shown in Figure 14 by \citet{Davis_2020}.

Because the primary spin was unaffected during the two obliquity transitions in the migration simulation, 
we infer that spin/spin resonant coupling \citep{Seligman_2021} 
between the two bodies was not a contributing factor. 
Rotation of a torque free triaxial body can be described in terms of precession
and nutation \citep{Boue_2009,Fahnestock_2008,Agrusa_2021}.  We tentatively associate the first obliquity transition with nutation excitation. 
The precession and nutation frequencies can be calculated using analytical or semi-analytical theoretical methods \citep{Boue_2009,Fahnestock_2008,Agrusa_2021} 
and the location of resonances associated with them could in future be predicted. 

The migration simulation shown in  Figure \ref{fig:b1c}
illustrates that if the system migrates inward via BYORP, non-principal axis rotation can be excited due to spin-orbit
resonances that involve obliquity, libration,  and non-principal axis rotation.
Thus even if tidal dissipation is strong enough to damp the secondary's obliquity and non-principal
axis rotation, inward migration could re-excite the secondary's rotation state. 

We also explored a simulation
that forced the primary to spin down, as might occur due to the YORP effect.
The primary spin drift rate used was $\dot \omega_p = - 2 \times 10^5 t_g^{-2}$ and as in our migration
simulation, initial conditions were the end state of the tidal spin down simulation shown in Figure \ref{fig:b1}.
The spin state of the secondary was unaffected until the primary spin reached the
spin/spin resonance $\omega_p \sim 2 \omega_s$ which caused the binary to disrupt.
Only when the primary spin was low,  did we see evidence for spin-spin resonant coupling.
In this setting spin/spin resonant coupling \citep{Seligman_2021} could be important.

We explored a simulation similar to that shown in Figure \ref{fig:b1c} except
the secondary migrated outward (instead of inward) from  $a_B/R_p \approx  2 $ to 4 at  
a drift rate $\dot a_B =  5.0\times 10^{-6}$.  
This simulation showed a similar obliquity jump at $a_B/R_p \sim 2.1$ as 
the first transition in Figure \ref{fig:b1c}. This implies that the  resonance that excited
the long axis inclination is sensitive to the binary period.
Otherwise this simulation was dull.  After that transition,  we did not see 
obliquity or non-principal axis excitation.   Perhaps simulations that migrate outward 
more slowly (adiabatically) would reveal weaker resonant behavior. 


\section{Implications for BYORP drift}
\label{sec:BYORP}

In  section \ref{sec:sim1} our tidal spin down simulation showed that secondary obliquity and non-principal axis excitation can be long-lived within the 1:1 spin-orbit resonance.  
In section \ref{sec:mig} our migration simulation showed that obliquity, long axis inclination and
libration 
can be excited if the binary semi-major axis drifts inward due to BYORP.
In this section we explore effect of non-principal axis rotation
on the BYORP drift rate for a secondary that is within the 1:1 spin-orbit resonance.   

Prior calculations of the BYORP affect \citep{Cuk_2005,McMahon_2010b,Steinberg_2011,Scheirich_2021} assume
that the secondary is tidally locked (in the 1:1 spin synchronous state), undergoes principal axis
rotation, is at zero obliquity and lacks free libration.
To estimate the role of non-principal axis rotation in affecting the BYORP induced drift rate we compute the BYORP torque on the binary orbit using different secondary shape models that are comprised of triangular surface meshes.

The shapes of the secondaries in most binary systems are not well characterized.
An exception is with the Binary Near-Earth Asteroid (66391) 1999 KW4, with primary known as Moshup and secondary known as Squannit. Squannit's shape model was derived by \citet{Ostro_2006} and a BYORP generated drift rate was computed with this shape model by \citet{Scheirich_2021}.
We use the same Squannit shape model to compare our BYORP drift rate estimates to that computed by \citet{Scheirich_2021}.  
 
To estimate the torque from BYORP we neglect surface thermal inertia, following \citet{Rubincam_2000}, so that thermal radiation is re-emitted with no time lag. 
The reflected and thermally radiated components are assumed to be Lambertian so they are emitted with flux parallel to the local surface normal.  We ignore heat conduction and neglect shadows on the day-lit side. 
We ignore variations of albedo or emissivity on the surface. Secondary intersections of re-radiated energy are ignored.

The asteroid surface is described with a closed triangular oriented vertex-face mesh, similar to the calculations by \citet{Scheeres_2007b,McMahon_2010b,Steinberg_2011}.   We index each triangular facet with an integer $i$.  The $i$-th facet has area $S_i$, and unit normal $\hat {\bf n}_i$. 
The radiation force on the $i$-th triangular facet is computed as 
\begin{equation}
{\bf F}_i  = \begin{cases} 
 - \frac{2}{3} \frac{F_\odot}{c} {S_i} (\hat {\bf n}_i \cdot \hat {\bf s}_\odot) \hat {\bf n}_i & \mathrm{if \ } \hat {\bf n}_i \cdot \hat {\bf s}_\odot >0 \\
0 & \mathrm{otherwise}
\end{cases},
\end{equation}
where $F_\odot$ is the solar radiation flux and $c$ is the speed of light.
The direction of the Sun is given by unit vector $\hat {\bf s}_\odot$. 
The factor of 2/3 is consistent with Lambertian reflectance and emission.
Facets on the day side have
$\hat {\bf n}_i \cdot \hat {\bf s}_\odot >0$ 
and we assume that only these contribute to the radiation force.  
In terms of the radial distance $r_h$ from the binary to the Sun, 
    $F_\odot \equiv \frac{P_\odot}{r_h^2}$
where radiation pressure constant
\begin{equation}
P_\odot \equiv \frac{L_\odot}{4 \pi c} = 10^{17} \ {\rm kg\ m \ s}^{-2}  \end{equation}
and where $L_\odot$ is the solar luminosity.

The torque affecting the binary orbit from a single facet is 
\begin{equation}
 {\boldsymbol \tau}_{i,B} = 
\begin{cases} 
- \frac{2}{3} \frac{F_\odot}{c} {S_i} (\hat {\bf n}_i \cdot \hat {\bf s}_\odot) ( {\bf a}_B \times \hat {\bf n}_i)  
 & \mbox{if } \hat {\bf n}_i \cdot \hat {\bf s}_\odot >0  \\
 0 & \mbox{otherwise}
 \end{cases},
\end{equation}
where ${\bf a}_B$ is the secondary's radial vector from the binary center of mass.
We neglect the distance of the facet centroids from the secondary center of mass when computing the radiative torques.

The torque affecting the binary orbit is the sum of the torques from each facet. 
This torque is averaged over different binary positions in its orbit around the Sun, affecting the direction of ${\bf s}_\odot$, and
over the different positions in the binary mutual orbit and spin rotation of the secondary,
\begin{equation}
{\boldsymbol \tau}_{BY} 
= \frac{1}{T} \int_0^T dt\  \sum_{i} 
{\boldsymbol \tau}_{i,B} .
\end{equation}
To represent this average, 
time $T$ significantly exceeds binary and heliocentric orbit periods. In practice we sum over different orbital phases to compute the average. The average can also be done by integrating over the mean anomaly $M_h$ of the binary in its heliocentric orbit and the mean anomaly of the binary orbit $M_B$
\begin{equation}
{\boldsymbol \tau}_{BY} = 
\int_0^{2 \pi}\int_0^{2 \pi}
dM_h \ dM_B \ 
\sum_{i} 
{\boldsymbol \tau}_{i,B}.
\end{equation}

If $\hat {\bf l}_B$ is the binary orbit normal then the mutual semi-major axis drift rate is 
\begin{equation}
\dot a_B = \frac{2 
{\boldsymbol \tau}_{BY} \cdot \hat {\bf l}_B}{M_s n_B a_B}.  \label{eqn:dotaB}
\end{equation} 


BYORP is often described in terms of a dimensionless number (e.g., \citealt{McMahon_2010b}).
By summing over different orientations and face model facets,  we calculate the dimensionless number
\begin{equation}
B_{BY} \equiv
   ({\boldsymbol \tau}_{BY} \cdot \hat {\bf l}_B ) \frac{c}{F_\odot R_s^2 a_B} .
   \label{eqn:gBY}
\end{equation}
This is equivalent to letting $a_B=R_s=F_\odot/c=1$. 
For a binary in a non-circular heliocentric orbit, averaging of the solar flux would give 
$\bar F_\odot = \frac{P_\odot}{a_h^2 \sqrt{1 - e_h^2}}$ where $a_h$ and $e_h$ are the binary orbit's heliocentric semi-major axis and eccentricity.
With our dimensionless number,  $B_{BY}$,  equations  \ref{eqn:dotaB} and \ref{eqn:gBY} give
\begin{align}
    \dot a_B 
    & = \frac{2 P_\odot}{a_h^2 \sqrt{1 - e_h^2}} \frac{R_s^2}{M_s n_B}  g_{BY}.
    \label{eqn:dota2}
\end{align}
We compare our expression to 
equation 3 by \citet{Scheirich_2021}
\begin{align}
    \dot a_{B}  
    & = \frac{2 P_\odot}{a_h^2 \sqrt{1 -e_h^2}} \frac{R_{mean}^2}{M_s n_B} B.
\end{align}
Comparison between this equation and equation \ref{eqn:dota2} shows that our dimensionless $B_{BY}$ parameter of equation \ref{eqn:gBY}  is equivalent to the dimensionless BYORP $B$ parameter introduced by \citet{McMahon_2010b}.
Our parameter  $B_{BY} = \pi f_{BY} $ where $f_{BY}$ is the dimensionless parameter computed by \citet{Steinberg_2011}. 

\subsection{Shape models}
\label{sec:shape}

To explore how BYORP depends on surface shape we use 3 shape models that are triangular surface meshes. 
The three shape models are an icosphere, a coarse version of the Squannit shape model by \citet{Ostro_2006} and a perturbed version  of the same Squannit shape model.
With the Squannit shape model
by \citet{Ostro_2006},
\citet{Scheirich_2021} computed a  BYORP B-coefficient of $B = 2.082\times10^{-2}$ for the Moshup/Squannit binary system.  
To test our BYORP B-coefficient computation,  we compare our computed B-coefficient to that computed by \citet{Scheirich_2021}. 

The icosphere is an icosahedron that is subdivided twice.  Each triangular face of the icosahedron is divided into 4 triangular faces and then each of them is divided into 4 triangular faces giving a total of 320 triangular faces.  The vertex positions are then corrected so they  have distance 1 from the origin.

\begin{figure}
    \centering
        \includegraphics[width=2.2in]{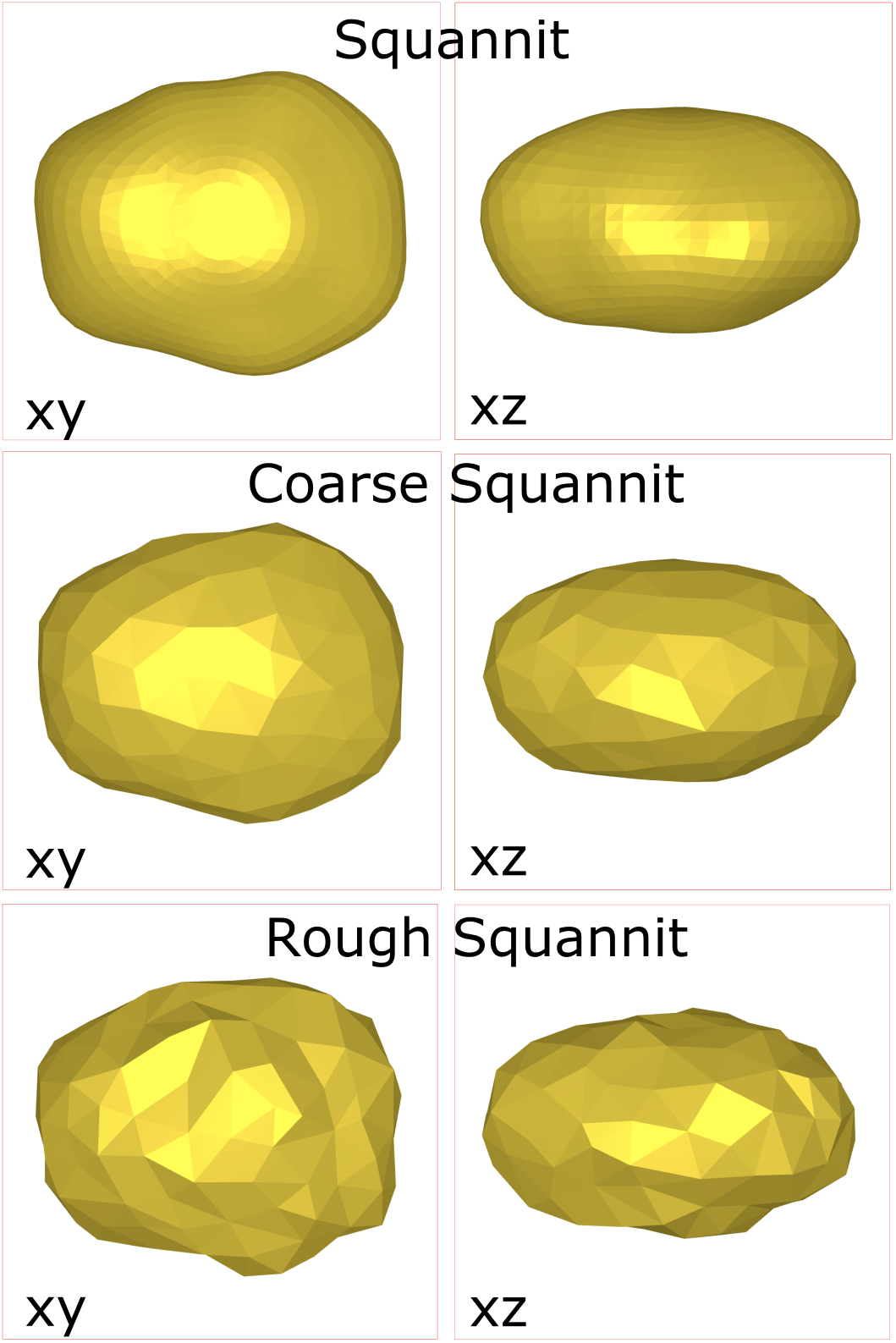}
    \caption{Triangular surface mesh shape model for Squannit by Ostro et al.\ (2006) is shown in the top two panels. The $xy$ plane is viewed on the left and the $xz$ plane is viewed on the right.  
    This shape model has 2292 faces.   The middle two panels shows a coarser version of the same shape model with 302 faces. 
    The bottom two panels show a model that is similar to the coarse one,   but each vertex position has been perturbed in a random direction to increase the surface roughness.  The red bounding squares have a length of 3 in units of the radius of the volume equivalent sphere.
    \label{fig:squannit_shape}}
\end{figure}

The Squannit shape model by \citet{Ostro_2006} contains 2292 faces is shown with $xy$ and $yz$ plane views in the top two panels of Figure \ref{fig:squannit_shape}.  
The red bounding squares in this Figure have a size of 3 in units of the radius of a volume-equivalent sphere.
To speed up our BYORP calculation we compute the BYORP drift rate for a similar but coarser triangular surface mesh with fewer faces.  We reduce the face number by merging short edges that are less than a specified length. We adjusted the specified minimum edge length to give a triangular surface mesh with about 300 faces.  
The coarser version of the Squannit shape model with 302 faces is shown in the middle panels of \ref{fig:squannit_shape}.
We checked that the results of our BYORP B-coefficient calculations are not sensitive to the number of faces used in our model (coefficients were less than a few percent different than those computed with  
a surface mesh that has twice the number of faces).

To create a rougher version of the Squannit shape model  we perturbed each vertex
of the coarse Squannit 302 face shape model by adding random values to the  $x,y$ and $z$ coordinates
of each vertex. 
The random values are taken from a uniform distribution spanning $[-\Delta, \Delta]$ where $\Delta =0.05$ in units of the radius of the volume-equivalent sphere.  The perturbed 302 face Squannit shape model is shown in bottom two panels of Figure \ref{fig:squannit_shape}.  The perturbations to the vertices 
make it rougher than the original and coarse Squannit shape models. 


\begin{figure}
    \centering
    \includegraphics[width=3.5in, trim = 15 0 0 0, clip]{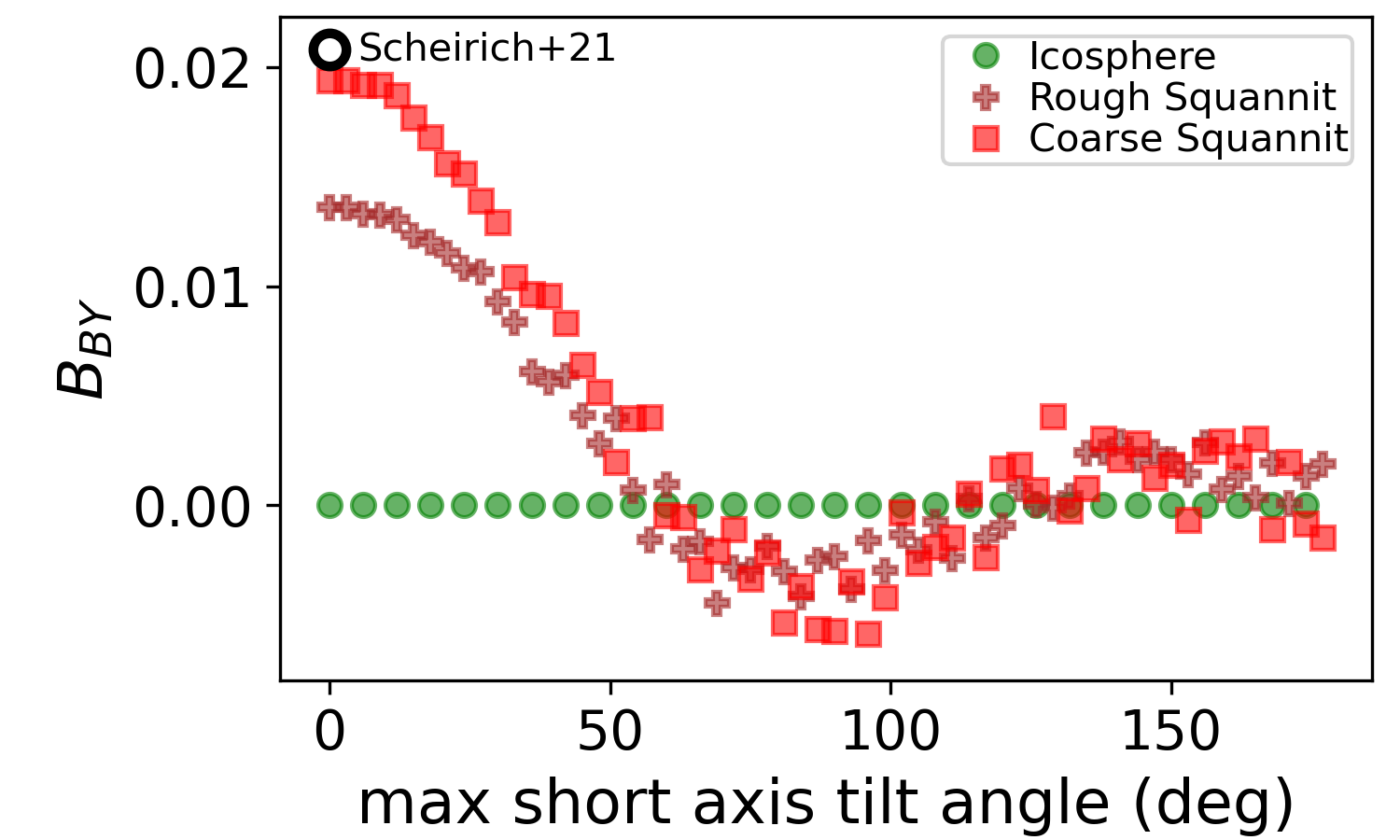}
    \caption{BYORP B-coefficients computed using two Squannit secondary shape models shown in the bottom
    four panels of Figure \ref{fig:squannit_shape}, as a function of the maximum tilt angle of the body's short axis from the orbit normal. 
    The body tilt is randomly chosen at each binary orbit orientation angle used in the torque averaging computation.
    The black circle is the B-coefficient computed by \citet{Scheirich_2021} based on the Squannit model by \citet{Ostro_2006}.
    The green circles are computed for an icosphere.  The red squares show the coarse Squannit shape model with 302 faces.  The brown crosses show a perturbed or rough Squannit shape model, also with 302 faces.
    The BYORP B-coefficient reverses sign when the short angle tilts by more than about $60^\circ$.
    \label{fig:tilt}}
\end{figure}

\subsection{BYORP B-coefficients computed with randomly chosen short axis tilt angles}
\label{sec:tilt}

Using the three shape models described in section \ref{sec:shape}, 
we numerically compute BYORP B-coefficients in two settings.
In our first set of calculations, 
we compute B-coefficients assuming a circular binary orbit and choosing orientations for the secondary at different positions in the orbit. In our second set of calculations (section \ref{sec:binBY}), we use the secondary orientations measured directly from our simulations to compute the B-coefficients.

In our mass-spring model simulations described in section \ref{sec:sims}, 
non-principal axis rotation of the secondary was present at low libration angle and was characterized by rotation about the secondary's long axis.   The secondary's long axis remains aligned with the direction to the primary, corresponding to $\phi_{lib,s} \approx 0$, however the short axis can tilt away from the orbit normal.  We described this rotation as the short axis tilt and it is measured from the angle between the body short axis and the orbit normal.  
We assume that the secondary angular rotation rate $\omega_s = n_B$ is equal to the 
binary orbit mean motion.  This condition places the secondary in the 1:1 spin synchronous resonance. We orient the secondary so that its long axis remains in the orbital plane and points  directly toward the primary so that the libration angle remains at zero.  At each binary orbit orientation, the secondary's short axis orientation is randomly chosen using a uniform distribution so that the angle between the short axis and the orbit normal ranges from 0 to a maximum of $\theta_{\rm max tilt}$.
Specifically, after rotating the secondary so that its long axis
is pointed toward the primary, the secondary is rotated about its long axis by an angle chosen from a uniform distribution within 
$ \{ -\theta_{\rm max tilt}, 
\theta_{\rm max tilt} \}$.

We assume a zero inclination circular orbit for the binary (about the Sun) and a circular mutual binary orbit.  
To numerically calculate the BYORP B-coefficient we first average over the Sun's possible directions using 36 equally spaced solar angles.    Then we average over the binary orbit using 144 equally spaced binary orbit orientations.
In each of these 144 possible orientations the short axis tilt is randomly chosen.

The BYORP B-coefficients computed using the three shape models shown in Figure \ref{fig:squannit_shape} and with randomly chosen short axis tilt angles are shown in Figure \ref{fig:tilt} as a function  $\theta_{\rm max tilt}$.  The green circles show that the B-coefficient computed for the icosphere is zero, as expected.    The coarse Squannit face model, shown with red squares, gives a BYORP B-coefficient of $B_{BY} = 0.02$ for $\theta_{\rm max tilt} = 0$ and this 
agrees with the value computed by \citet{Scheirich_2021} for Squannit that is shown with a black circle.  
For both Squannit and rough Squannit shape models, the BYORP B-coefficient drops to zero at a maximum tilt angle of about $\theta_{\rm max tilt} \sim 60^\circ$.
This gives a quantitative estimate for how much non-principal axis rotation within the 1:1 spin-orbit resonance is required to reduce BYORP drift.

The rough or perturbed Squannit shape model gives lower values of
the BYORP B-coefficient than the Squannit shape model.   This is opposite to the expected trend.
 Usually rougher surface models have higher YORP and BYORP coefficients  \citep{Steinberg_2011}.  
 By computing BYORP B-coefficients for perturbed ellipsoids we
 confirmed that  rougher surfaces tend to have larger B-coefficients. 
The specific shape of the Squannit shape model must be responsible for its high B-coefficient.  
The ratio of B-coefficients for Squannit and perturbed Squannit shape models we computed
at $\theta_{\rm maxtilt} =0$ is lower than 2.
In the Moshup/Squannit system, 
\citet{Scheirich_2021} measured a drift rate about 8 times lower than predicted.
A rougher surface for Squannit does not
give a large enough difference in the B-coefficient to fully account for its low measured B-coefficient (assuming
principal axis rotation).   In contrast, the BYORP B-coefficient can be negligible 
if there is non-principal axis rotation with 
$\theta_{\rm maxtillt} \sim 60^\circ$.

\begin{figure}
    \centering
    \includegraphics[width=3.5in, trim = 20 0 0 0, clip]{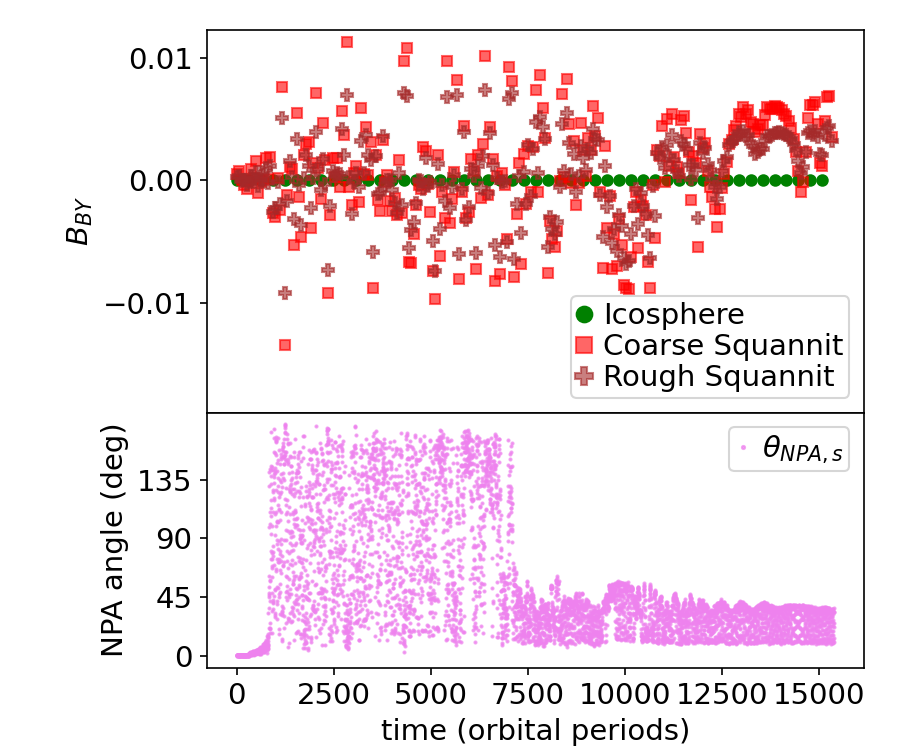}
    \caption{BYORP-B coefficient is computed using secondary and binary orientations from the
    tidal spin-down simulation shown in Figure \ref{fig:b1} and is shown 
    in the top panel as a function of time in orbital periods.   The B-coefficients were computed
    using our three shape models (see Figure \ref{fig:squannit_shape}). 
    The red squares are the coarse Squannit shape model, 
    the brown pluses are the perturbed (or rougher) 
    Squannit shape model. 
    The green circles show an icosphere shape model.
    The icosphere has a B-coefficient of zero, as expected. 
    The secondary's non-principal axis angle $\theta_{\rm NPA,s}$ is shown in the bottom panel.
      The BYORP B-coefficients for the Squannit shape models frequently reverse in sign. Only
      after the NPA angle has dropped below $45^\circ$ near the end of the simulation does
      the BYORP B-coefficient remain positive.  The rougher Squannit shape model has
      B-coefficients similar to but lower amplitude than the coarse Squannit shape model.
      Both Squannit shape modes have BYORP B-coefficients lower than predicted for a
      tidally locked body undergoing principal axis rotation, at zero obliquity and with no free libration. 
      The conventionally predicted B-coefficients are 0.02 for the coarse Squannit model 
      and 0.014 for the rough Squannit model.  This figure shows that long-lived non-principal axis
      rotation within the 1:1 spin-orbit resonance
      can cause a reduction in the amplitude of the BYORP B-coefficient and reversals in 
      the direction of BYORP effect drift.    \label{fig:BY_b1}}
\end{figure}

\begin{figure}
    \centering
    \includegraphics[width=3.5in, trim = 20 0 0 0, clip]{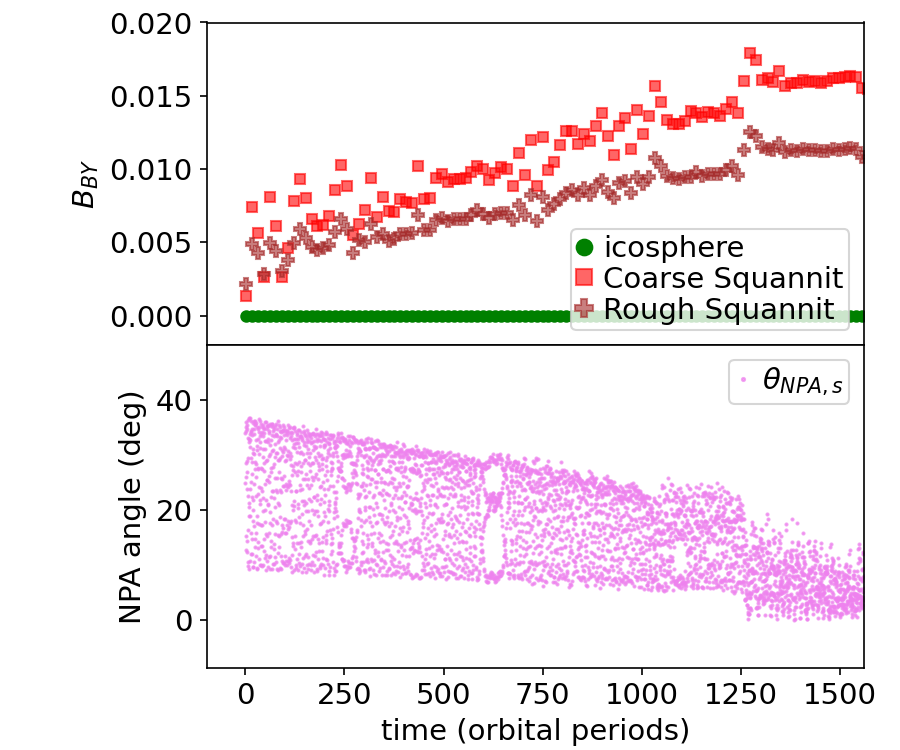}
    \caption{Similar to Figure \ref{fig:BY_b1} except the BYORP B coefficient is 
    computed using the migration simulation shown in Figure \ref{fig:b1c}.    As the non-principal axis rotation angle
    drops, the BYORP B-coefficients for the Squannit shape models increase.  
    \label{fig:BY_b1cc}}
\end{figure}

\subsection{BYORP B-coefficients computed using the mass-spring model numerical simulations}
\label{sec:binBY}

We find the secondary orientation at each simulation output in our mass-spring model simulations. We use the vector between secondary and primary, binary orbit normal and body orientation at each simulation output to compute the BYORP B-coefficient for three shape models, two of which are shown in Figure \ref{fig:squannit_shape}. 

The secondary orientation (computed with a quaternion from a simulation output as described in section \ref{sec:simout}) was used to rotate the shape model vertices.
The rotated shape model vertices, face normals and face areas were then used to compute the radiative torque on each facet using binary orientation vector and orbit normal from the concurrent simulation output.
At each simulation output, 18 solar positions were used to compute an average BYORP torque.   100 consecutive simulation outputs, spanning about 6 binary orbit periods, were then combined for an average of the BYORP induced torque at a particular moment during the simulation.  We note that the axis ratios of the Squannit shape model ($b_s/a_s = 0.762$, $c_s/a_s = 0.576$;  \citealt{Ostro_2006}) are similar to but not exactly the same as those of the secondary in the simulation (0.773, 0.698).  

To check our numerical routines we ran a short simulation of a secondary in the spin synchronous state,  zero obliquity and in a principal axis rotation rate.  We  measured the BYORP B-coefficient in this short simulation.  The measured B-coefficient values for each shape model were consistent with those computed at $\theta_{\rm max tilt} = 0$, and shown in Figure \ref{fig:tilt}, for the same shape models, as expected.  The B-coefficient computed
for the coarse Squannit shape model again agreed with that computed by \citet{Scheirich_2021}, $B_{BY} \approx 0.02$.

Figure \ref{fig:BY_b1} shows the BYORP  B-coefficients computed for our three shape models from the tidal spin-down simulation shown in Figure \ref{fig:b1} as a function of time and Figure \ref{fig:BY_b1cc}
similarly shows the BYORP B-coefficients but computed for the migration simulation shown in Figure \ref{fig:b1c}.
In these figures, green circles refer to the icosphere shape model, red squares refer to the 302 face coarse Squannit shape model and brown pluses refer to the rougher Squannit shape model.  In both figures the top panels show the B-coefficients and the bottom panels show the non-principal axis angle $\theta_{\rm NPA,s}$.   

Figure \ref{fig:BY_b1} shows that 
the BYORP B-coefficients are lower than those computed for principal axis rotation state ($B_{BY} \approx 0.020$ for the Squannit shape model and $B_{BY} \approx 0.014$ for the perturbed or rough Squannit shape model) 
throughout the tidal-spin down simulation. 
The sign of the B-coefficient frequently reverses which implies that the cumulative B-coefficient averages to zero.
Not until the end of the simulation do the reversals cease and the BYORP B-coefficient remains above zero.  At the end of the simulation the average B-coefficient is about 4 times lower than that in a principal axis rotation state.  The simulation ran for 15000 orbital periods but the 1:1 spin-orbit resonance was reached after only about 1000 orbital  periods.  The BYORP torque was low for a duration an order of magnitude longer than the spin down time.  We infer that long-lived non-principal axis rotation within the 1:1 spin-orbit resonance may prolong the lifetime of the binary by reducing BYORP effect orbital drift.

In  Figure \ref{fig:BY_b1} the BYORP B-coefficient for the Squannit shape models showed reversals
even for non-principal axis angles $\theta_{\rm NPA,s}$ as low as $45^\circ$. This is smaller
than the maximum short axis tilt angle that gave a B-coefficient of zero in Figure \ref{fig:tilt}.
In the computations described in section \ref{sec:tilt}, we assumed a uniform distribution
for the short axis tilt and neglected variations in libration and long axis inclination. 
This may not be a good approximation for the orientation angle distributions exhibited by the
secondary in the numerical simulations.   A comparison between the red squares
and brown pluses in Figure \ref{fig:BY_b1} shows that the rough Squannit model has BYORP B-coefficients 
similar to but lower in amplitude than that of the coarse Squannit model, as we found previously in section \ref{sec:tilt}.

The migration simulation shown in Figures \ref{fig:b1c} and \ref{fig:BY_b1cc} has initial condition starting where the tidal-spin down simulation ended.  In Figure \ref{fig:BY_b1cc}, 
as the NPA angle decreases, the amplitude of the BYORP B-coefficient slowly increases until it nears
the value expected for a tidally locked body undergoing principal axis rotation (0.02 for the Squannit shape model).    
This simulation exhibited jumps in obliquity and long axis inclination angle.   However the BYORP B-coefficients
slowly increase throughout the simulation (until right before the binary is disrupted) 
and more closely follow the NPA angle than the obliquity
or long axis tilt variations.    We suspect that the time averaged BYORP B-coefficient is more sensitive to the 
NPA angle variations than to the libration amplitude, or the long axis tilt and obliquity variations.

We have presented an estimate of the BYORP torque as a function of time using two spin-dynamics simulations and two shape models.  Future studies can
improve upon and extend this study by carrying out longer duration simulations (more orbits) with more slower tidal evolution or migration and 
with better resolved secondary shapes and internal structure.   
Future BYORP B-coefficients computed using spin dynamics simulations could explore how  
the sensitivity of the BYORP torque to the extent of non-principal axis rotation depends on body shape, internal structure (e.g., homogeneity) and surface properties such as albedo, reflectively, emissivity and thermal inertia. 

\section{Summary and Discussion}
\label{sec:sum}

Using our viscoelastic mass/spring model N-body code,  we have carried out long timescale
($10^4$ binary orbital period or about 15 year long) simulations of a tidally evolving asteroid binary system similar to the NEA (65803) Didymos binary system. 
We begin a simulation with secondary outside of tidal lock, undergoing principal axis rotation  and at low obliquity.  Due to attitude instability when crossing into the 1:1 spin-orbit resonance, the secondary's obliquity is excited and it enters a non-principal axis rotation state \citep{wisdom87}.   We find that the obliquity and non-principal axis rotation damping time exceeds the time to enter the spin synchronous state by about an order of magnitude, so secondary obliquity and non-principal axis rotation can be long-lived in NEA binary secondaries, perhaps even lasting millions of years.  

While in the 1:1 spin-orbit resonance and in a non-principal axis rotation state,   
the secondary's long axis librates about the direction of the primary
but the secondary's short axis can exhibit wide excursions away from the orbit normal.  
In such a state, it may be difficult to identify non-principal axis rotation from 
a light curve, and yet, the secondary could be correctly identified as residing in the 1:1 spin-orbit resonance.
We propose that some binary asteroids that have been classified as having synchronous secondaries from their light curves may reside in the 1:1 spin-orbit resonance but may simultaneously reside in a complex non-principal axis rotation state.

Our numerical simulations confirm the work by \citep{Naidu_2015} who emphasized that non-principal axis rotation may be slowly damped.  
Long-lived non-principal axis rotation was also previously seen in some simulations of a tidally evolving Deimos \citep{quillen20_phobos}.

If binary secondaries are observed to undergo principal axis rotation, have near-zero obliquity and low amplitudes of free libration, then they would require even higher levels of tidal dissipation than previously estimated to reach their tidally locked states.
Lower required values of $\mu Q$ (shear modulus times tidal dissipation parameter) would exacerbate the tension between viscoelastic material properties estimated for single and binary asteroids.  
In contrast, if the secondaries are observed to be in non-principal axis rotation states, then higher values
of $\mu Q$ would be possible and the discrepancy between estimated single and binary asteroids material properties may be ameliorated.

To mimic the effect of BYORP induced migration, we carried out a numerical simulation of a binary system initially in a low obliquity, low principal axis rotation and spin-synchronous rotation state.
Spin orbit resonances, crossed during migration, can excite obliquity, libration and non-principal axis rotation.  
Even if tidal dissipation is strong enough to damp the secondary's obliquity and non-principal axis rotation, 
BYORP induced migration could re-excite the secondary's rotation state. 
We  saw resonant excitation of obliquity and non-principal axis rotation but only during inward migration. 
Impacts on the secondary or surface movements on it might also re-excite its rotation state. 

Previous BYORP calculations have necessarily assumed zero secondary obliquity,  principal axis rotation
and low libration amplitude within the 1:1 spin-orbit resonance.  
An expectation is that when the secondary tumbles, 
the cumulative BYORP torque will average to zero \citep{Cuk_2005,Naidu_2015}.  
However, our simulations show  an intermediate rotation state where the secondary remains within the 1:1 spin-orbit resonance and simultaneously exhibits chaotic non-principal axis variations.  
Using a shape model \citep{Ostro_2006} for Squannit, the secondary of the Moshup system (NEA (66391) 1999 KW4),  we estimate the BYORP B-coefficient.  We find that the BYORP torque is dependent upon the extent of minor axis tilt within the 1:1 spin synchronous state.  
There is a range of possible BYORP torque values that is 
dependent upon the extent of non-principal axis rotation.

We use secondary body orientations measured in our mass-spring model simulations 
and Squannit shape models  to compute the BYORP B-coefficient.   The B-coefficient is reduced
compared to that computed for the tidally locked, zero obliquity and principal axis rotation state.
The B-coefficient shows reversals in sign for an extended period of time after the spin-synchronous state is entered. 
Only when the non-principal axis angle drops below about $45^\circ$  does the B-coefficient and associated BYORP effect torque remain unidirectional.
The lifetime of the binary would be extended due to non-principal axis rotation.  However, we find that non-principal axis rotation excitation due to inward migration may not be strong enough to entirely prevent migration by the BYORP effect.  Once the secondary non-principal axis angle drops below about $45^\circ$,  and 
unless non-principal axis rotation in the secondary is re-excited by impacts or mass-movements,  the BYORP induced orbital migration rate can increase until it is as rapid as estimated previously for the principal axis rotation state \citep{Cuk_2005}. 

Non-principal axis rotation of Squannit, the secondary in the Moshup system, could help account for its low BYORP measured drift rate, which was about 7 times lower than that predicted from the secondary shape model \citep{Scheirich_2021}.   To a lesser extent, a shape model with a rougher surface would 
also give a lower B-coefficient.  As discussed by \citet{Scheirich_2021}, the discrepancy between predicted and measured BYORP drift rate may simply be due to uncertainties in the secondary's topography.

The DART and Hera missions to Didymos, on-going photometric surveys and radar studies can confirm or rule out the possibility that binary asteroid secondaries are in complex rotation states.  
If binary secondaries are found in complex rotation states, then tidal dissipation need not be as high as that required to balance the conventionally estimated BYORP torque, as previously proposed  by \citet{Jacobson_2011}.   
As the rotation state affects the BYORP induced orbital drift rate, binary population evolution models \citep{Jacobson_2016} could be modified to take into account the role and evolution of the secondary rotation state.
Long duration spin dynamics simulations could be modified to take into account YORP and BYORP effect induced torques.  Meteroid impacts and mass movements can perturb the spin state and body shape, and modify surface properties such as albedo, reflectively, emissivity and thermal inertia, which affect the radiative torque.
As a consequence, binary asteroid population evolution models could take into account these additional processes. 

\vskip 0.3 truein

\vskip 1 truein 
\textbf{Acknowledgements}

This material is based upon work supported in part by NASA grants 80NSSC21K0143 and
80NSSC17K0771, and National Science Foundation Grant No. PHY-1757062.

We thank Patrick Michel for correspondence that helped us improve the manuscript.  

A repository for code developed and used in this paper is at 
\url{https://github.com/aquillen/asteroid_binary}.

\vskip 0.3 truein
 

\bibliographystyle{elsarticle-harv}
\bibliography{refs_bin}

\end{document}